\documentclass[%
 aip,
 amsmath,amssymb,
 reprint,
]{revtex4-1}

\usepackage{graphicx}
\usepackage{dcolumn}
\usepackage{bm}
\usepackage{subcaption}
\usepackage{siunitx}
\usepackage{float}
\usepackage{graphicx,booktabs,array}
\usepackage[utf8]{inputenc}
\usepackage[T1]{fontenc}
\usepackage{mathptmx}
\usepackage[english]{babel}

\begin{document}

\preprint{AIP/123-QED}

\title{Evolution of the self-injection process in the transition of an LWFA from self-modulation to blowout regime}

\author{Prabhat Kumar}
\affiliation{Department of Applied Mathematics and Statistics, Stony Brook University, Stony Brook, New York 11794, USA.}
\author{Kwangmin Yu}%
\affiliation{Computational Science Initiative, Brookhaven National Laboratory, Upton,
 New York 11973, USA.}%
\author{Rafal Zgadzaj}
\affiliation{Department of Physics, University of Texas at Austin, Austin, Texas 78712, USA.}
\author{Michael Downer}
\affiliation{Department of Physics, University of Texas at Austin, Austin, Texas 78712, USA.}
\author{Irina Petrushina}%
\affiliation{Department of Physics and Astronomy, Stony Brook University, Stony Brook 11794 , NY, USA.}
\author{Roman Samulyak}
\email{roman.samulyak@stonybrook.edu}
\affiliation{Department of Applied Mathematics and Statistics, Stony Brook University, Stony Brook, New York 11794, USA.}
\affiliation{Computational Science Initiative, Brookhaven National Laboratory, Upton, New York 11973, USA.}
\author{Vladimir Litvinenko}
\affiliation{Department of Physics and Astronomy, Stony Brook University, Stony Brook 11794 , NY, USA.}
\author{Navid Vafaei-Najafabadi}
\affiliation{Department of Physics and Astronomy, Stony Brook University, Stony Brook 11794 , NY, USA.}
\date{\today}

\begin{abstract}
	
Long wavelength infrared (LWIR) laser driven plasma wakefield accelerators are investigated
here in the self-modulated laser wakefield acceleration (SM-LWFA) and blowout regimes using
3D Particle-in-Cell simulations. The simulation results show that in SM-LWFA regime, self-injection arises with wave breaking, whereas in the blowout regime, self-injection is not observed under the simulation conditions. The wave breaking process in SM-LWFA regime occurs at a field strength that is significantly below the 1D wave-breaking threshold. This process intensifies at higher laser power and plasma density and is suppressed at low plasma densities ($\leq 1\times10^{17}$ $cm^{-3}$ here). The produced electrons show spatial modulations with a period matching that of the laser wavelength, which is a clear signature of direct laser acceleration (DLA).

\end{abstract}

                             

\maketitle


\section{\label{Intro}Introduction}

Laser driven plasma accelerators have made significant advances in recent years, now achieving electron beams of multi-GeV energy, few percent energy spread, with charge above 100 pC, and normalized transverse emittance as small as 0.1 $\pi$ mm mrad in a cm-scale plasma\cite{Mangles2017,Hooker:2014,Gonsalves2019,Leemans2006}. Such high energy electron beams are capable of driving particle and coherent radiation sources, with applications in material science, chemistry, and medicine. However, these applications place stringent restrictions on the quality of the electron beams, in terms of energy spread, beam charge, emittance, and shot-to-shot instabilities. Maintaining the quality of the beams requires the ability to control and manipulate the injection process. Strong progress in laser wakefield acceleration (LWFA) research has been achieved largely by using ultrashort, high power driver lasers operating at near infra-red wavelengths of 0.8-1 $\mu m$. With these laser parameters, the accelerating plasma structures are relatively small, with dimensions on the order of 10 to few tens of micrometers, leading to very high accelerating gradients but difficulties with precise injection, control and visualization, making free electron lasers (FEL) and high energy Physics (HEP) applications difficult. The choice of ~1 $\mu m$ driver wavelengths has been motivated by available technology and is not necessarily the optimal choice.  Recent advances in CO\textsubscript{2} laser technology \cite{Polyanskiy:15,Polyanskiy2020} has lead to the generation of terawatt-class peak power lasers in the long-wavelength infrared (LWIR) spectral domain. This has opened the possibility of exploring longer wavelength drivers, which thanks to favorable wavelength scaling, are more efficient at driving plasma waves. LWIR could ultimately drive fully blown out plasma bubbles with dimensions of several hundreds of microns, in plasma densities on the order of $10^{16} \text{ cm}^{-3}$, relaxing the conditions for external injection into the accelerating phase of the bubble, leading to extremely small emittances and energy spreads \cite{PhysRevLett.92.054801}. 

The CO\textsubscript{2} laser parameters ($2.0$ ps pulse duration, $9.2 \mu m$ wavelength, and $2-5$ TW peak power) currently generated by the Accelerator Test Facility (ATF) of Brookhaven National Laboratory (BNL) \cite{Polyanskiy2020} has allowed the exploration of the self-modulated laser wakefield acceleration (SM-LWFA) in a previously inaccessible parameter regime and at densities between $\sim 10^{17} - 10^{18} cm^{-3}$. The first experimental observation of self-modulated wakes at these densities was reported \cite{Welch:17} and analyzed using Particle-in-Cell simulations where ionization was shown to play an important role in reproducing experimentally observed Stokes/anti-Stokes sidebands of symmetric amplitudes \cite{Kumar2018,Kumar2019}. 

In SMLWFA, relativistic plasma waves can reach wave-breaking limits and drive background plasma electrons to reach longitudinal velocity higher than the phase velocity of the plasma wave within the wake\cite{Esarey1996}. This so-called self-injection process traps and accelerates electrons to multi-MeV energies. Self-injection in this regime has been shown to be aided via coupling with Raman backscattering (RBS), thermal and 3D effects, and can occur below theoretically predicted wave-breaking limit\cite{Esarey1998}. Furthermore, since the laser pulse overlaps several plasma wakes, self-injected electrons also gain energy from the transverse fields of the laser pulse through the direct laser acceleration (DLA) mechanism\cite{ShawDLA,ShawDLA1,Shaw_2016,Zhang2015,Pukhov2003}. Contribution from different energy gain mechanisms depend on several laser and plasma parameters. At sufficiently high energies, laser undergoes relativistic self-channeling. The electrons trapped in the channel are overlapped by a significant portion of the laser pulse and gain longitudinal momentum through DLA process, producing multi-MeV energy electrons\cite{Borghesi1997,Gahn1999}. Although the quality of the electrons generated in the SM-LWFA \cite{Antonsen1993,Ting_1997,Chen_2000} makes them unsuitable to FEL and HEP applications (due to relatively low energies, large emittance, large energy spread), it has recently seen renewed interest as a strong betatron x-ray source \cite{Albert,Lemos_2016,Albert_2018}. 

LWIR laser-driven LWFA in the blowout and bubble regime is the most promising. ATF is poised to produce sub-picosecond pulses with powers $\sim 10-30$ TW in the near future, which will be capable of reaching the blowout regime and bubbles with sizes $\sim 100 $ microns or more \cite{Igor}. LWIR laser driven wakefield accelerators with external injection experiments must be designed to produce clean plasma bubbles by making sure that laser pulse does not overlap the accelerating phase of the plasma wake, and wakefield amplitudes do not grow to wave breaking limits triggering the process of self-injection. 

In this paper, CO$_2$ laser driven wakefield acceleration has been studied in the self-modulated regime for a variety of laser and plasma parameters motivated by experiment AE-93 being conducted at ATF through Particle-in-Cell (PIC) simulations. 3D simulations are performed using PIC code SPACE, which contains solvers for Vlasov-Maxwell and Vlasov-Poison equations with a focus on atomic physics transformations. The process of self-injection and acceleration will be examined and the plasma density threshold for self-injection process to begin at currently available laser parameters will be determined. Simulations with near future laser parameters will be presented to predict transition from self-modulated to highly promising blowout regime. Ideal plasma and laser parameters will be presented to achieve fully blown out bubbles without self-injected electrons.

\section{\label{Methods}Methods and Algorithms}
Simulations were performed using the 3D, Particle-in-Cell code SPACE, developed at Stony Brook University and Brookhaven National Laboratory. It is a parallel, relativistic PIC code with novel algorithms for atomic physics transformations induced by high energy laser-plasma and beam-plasma interactions\cite{Yu2015}[add SPACE CPC]. The electromagnetic module of SPACE utilizes Yee's FDTD method\cite{Yee1966} for solving field equations and Boris-Vay pusher\cite{Boris1971,Vay} for advancing macroparticles. Tunneling ionization algorithm based on ADK formulation\cite{Ammosov1986} has been implemented in code SPACE for laser-induced ionization\cite{Kumar2019}. A novelty of the code is its ability to compute ionization and recombination rates on the grid and transfer them to particles rather than using a Monte-Carlo based procedure used in several PIC codes. SPACE code has been parallelized using hybrid MPI-OpenMP approach and has been shown to scale up-to hundreds of processors for complex three-dimensional laser-plasma interaction problems. SPACE has been used for the simulation of several beam-plasma and laser-plasma experiments at Fermi Lab and Brookhaven National Lab\cite{Yu2017,JunCEC,Kumar2018,Kumar2019,Kumar2019a}. 
 
\begin{figure*}
	\includegraphics[width=0.8\linewidth]{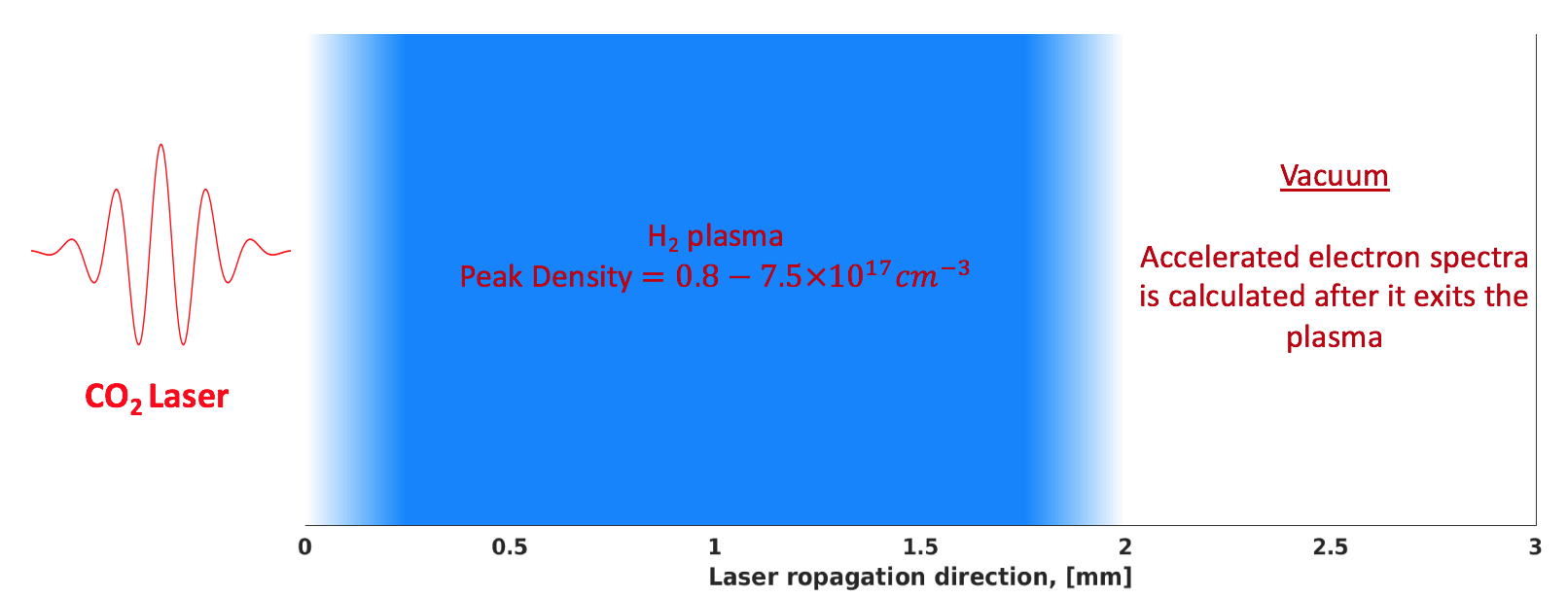}
	\caption{  \raggedright simulation set up}
	\label{set-up}
\end{figure*}

All the simulations were performed in a 3D Cartesian geometry. Computational box has a transverse size of 600 micrometers and a longitudinal size of 3-5 millimeters, with transverse resolution $dx = dy = 2.0\times 10^{-6}$ $m$ and longitudinal resolution $dz = 0.5\times 10^{-6}$ $m$. Simulations use a minimum of 32 macroparticles per cell.  Numerical convergence studies confirmed that this resolution is sufficient for the study of targeted problems. The plasma wave driver laser beam enters the simulation box from the left plane. It is assumed to be linearly polarized and has Gaussian longitudinal and transverse profiles. Computational domain is filled with uniformly distributed hydrogen gas with linear ramps ramps as shown in Figure \ref{set-up}. ADK tunneling ionization algorithm as is used to ionize the neutral and create plasma. Ions are assumed to be mobile. Laser and plasma parameters, and interaction length vary for different simulation cases and will be specified in section \ref{Results}.   


\section{\label{Results}Simulation Results and Discussion}
\subsection{SM-LWFA driven by CO$_2$ laser}
When a high-intensity laser pulse with a pulse duration larger than several plasma wavelengths propagates through an underdense plasma, it can self-focus and can also generate plasma waves through the Raman forward scattering (RFS) instability. For self focusing to occur, laser peak power (P) should be greater than the critical power, $(P_c)$, required for relativistic optical guiding of a long laser beam.  Here, 
\begin{equation}
P_c \approx 17\left(\frac{\lambda_p}{\lambda}\right)^2GW.
\label{P_c}
\end{equation}
In this so-called self-modulated laser wakefield acceleration (SM-LWFA) regime, some self-guiding of the laser pulse occurs due to relativistic modification of the plasma refractive index which maintains the laser pulse at high normalized vector potential for several Rayleigh lengths\cite{Esarey1996,Leemans1996}. In this process of interaction the laser pulse induces self-modulation that manifests itself as Stokes and anti-Stokes waves \cite{Esarey1996}. The frequency and wavenumber of these waves are given by $(\omega_0 - n\omega_p, k_0 - nk_p)$ and $(\omega_0 + n\omega_p, k_0 + nk_p)$ respectively, where $(\omega_0, k_0)$ is the frequency and wavenumber of the laser and $(\omega_p, k_p)$ is that of the plasma. Here, $n$ stands for  the  harmonic  number  of  the  RFS  sidebands.

The CO$_2$ laser parameters currently delivered by the ATF has allowed the exploration of the self-modulated regime driven by long-wavelength infrared (LWIR) lasers. The first experimental observation of self-modulated wakes in the density range between $\sim 10^{17} - 10^{18}$ $cm^{-3}$ were reported \cite{Welch:17,DPP19} and confirmed via 3D PIC simulations\cite{Kumar2019,Yan2018}. Here, we will present results of simulations motivated by the parameters used in experiment AE-93 at ATF.   

\begin{table}[h]
	\caption{Laser Parameters for Self-injection Study }
	\begin{ruledtabular}
		\begin{tabular}{cc}
			\textrm{Parameters}&
			\textrm{Value}\\
			\colrule
			Wavelength & $9.2 \mu  m$  \\
			Beam Waist & $20.0 \mu m$ \\
			Duration(FWHM) & $2 ps$ \\
			Energy & $4.0 J$  \\
		\end{tabular}
	\end{ruledtabular}
	\label{table2}
\end{table}

\begin{figure*}
	\includegraphics[width=\linewidth]{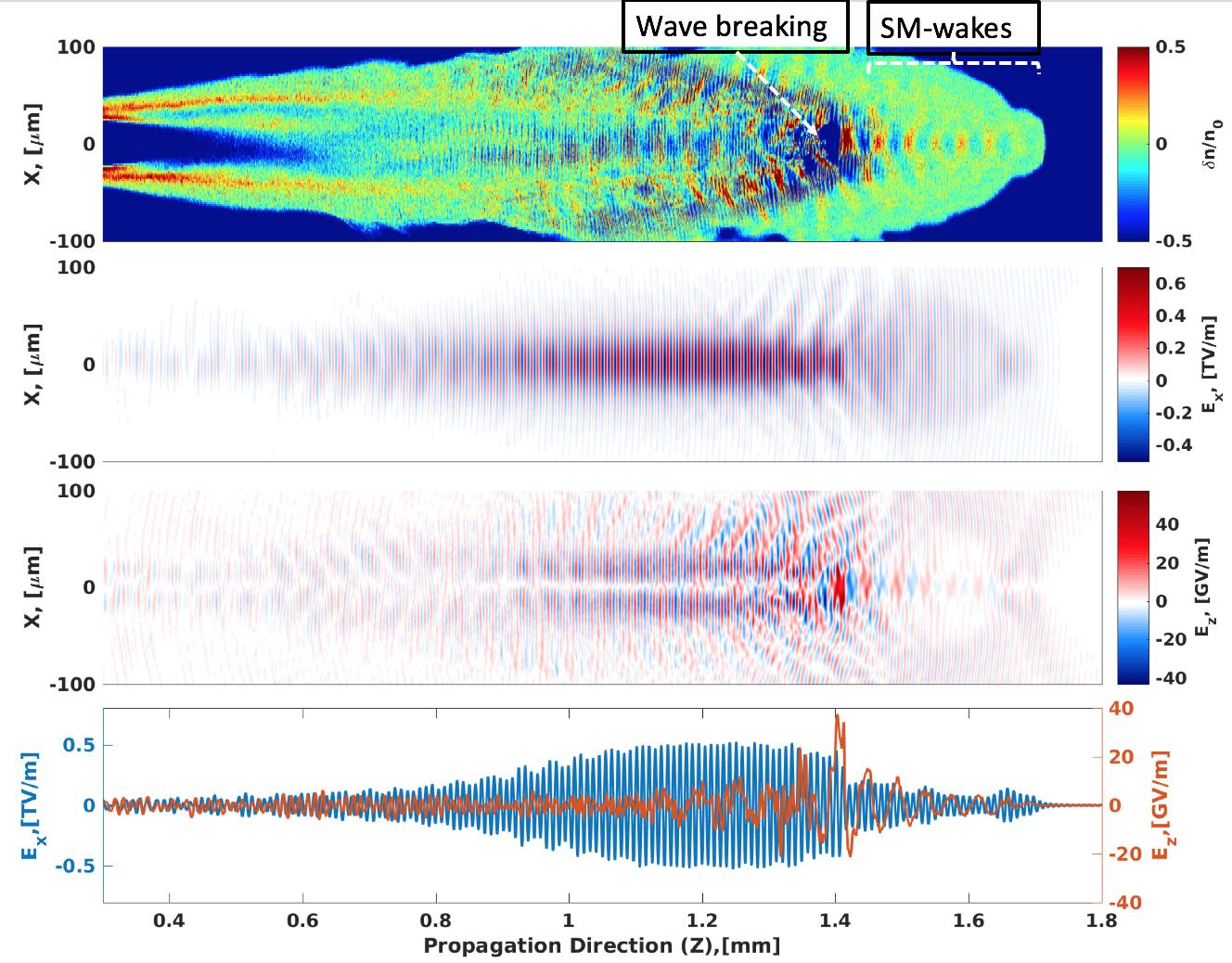}
	\caption{  \raggedright Structure of the self-modulated wakes and corresponding transverse and longitudinal fields, $(E_x$ and $E_z)$ respectively, are shown for plasma density $(n_e) = 5.0\times 10^{17}$ $cm^{-3}$. Bottom plot shows the axial line outs of these fields. Linearly polarized CO$_2$ laser is propagating from left to right. $X$ is the direction of linear polarization of the laser pulse.}
	\label{5e17chf}
\end{figure*}

We start with a 3D simulation of the interaction of a linearly polarized CO$_2$ laser pulse having parameters present in table \ref{table2} with hydrogen plasma. Neutral hydrogen gas is uniformly distributed in a $2.0$ $mm \times 600 $ $\mu m \times 600$ $\mu m$ simulation box with a $250$ $\mu m$ ramp at the entrance of the laser. Laser pulse, injected at the left boundary of the simulation box, ionizes the hydrogen gas giving a peak plasma density of $5.0 \times 10^{17}$ $cm^{-3}$. As the laser pulse traverses the simulation box from left to right, it undergoes self-modulation instability and Raman Forward Scattering instability creating self-modulated relativistic plasma wakes. The structure of the wake after a propagation of $1.8$ $mm$ is shown in the top row of figure \ref{5e17chf}. Clear self-modulated wakes can be seen in the front of the laser pulse. once the plasma
wake amplitude reaches wave breaking, it undergoes relativistic self-channeling. Transverse and longitudinal fields are shown in the two middle rows of figure \ref{5e17chf}. Axial line out of the fields (bottom row of figure \ref{5e17chf}) shows that the peak longitudinal field close to $40$ $GV/m$ is generated by the interaction.

\begin{figure*}
	\includegraphics[width=0.8\linewidth]{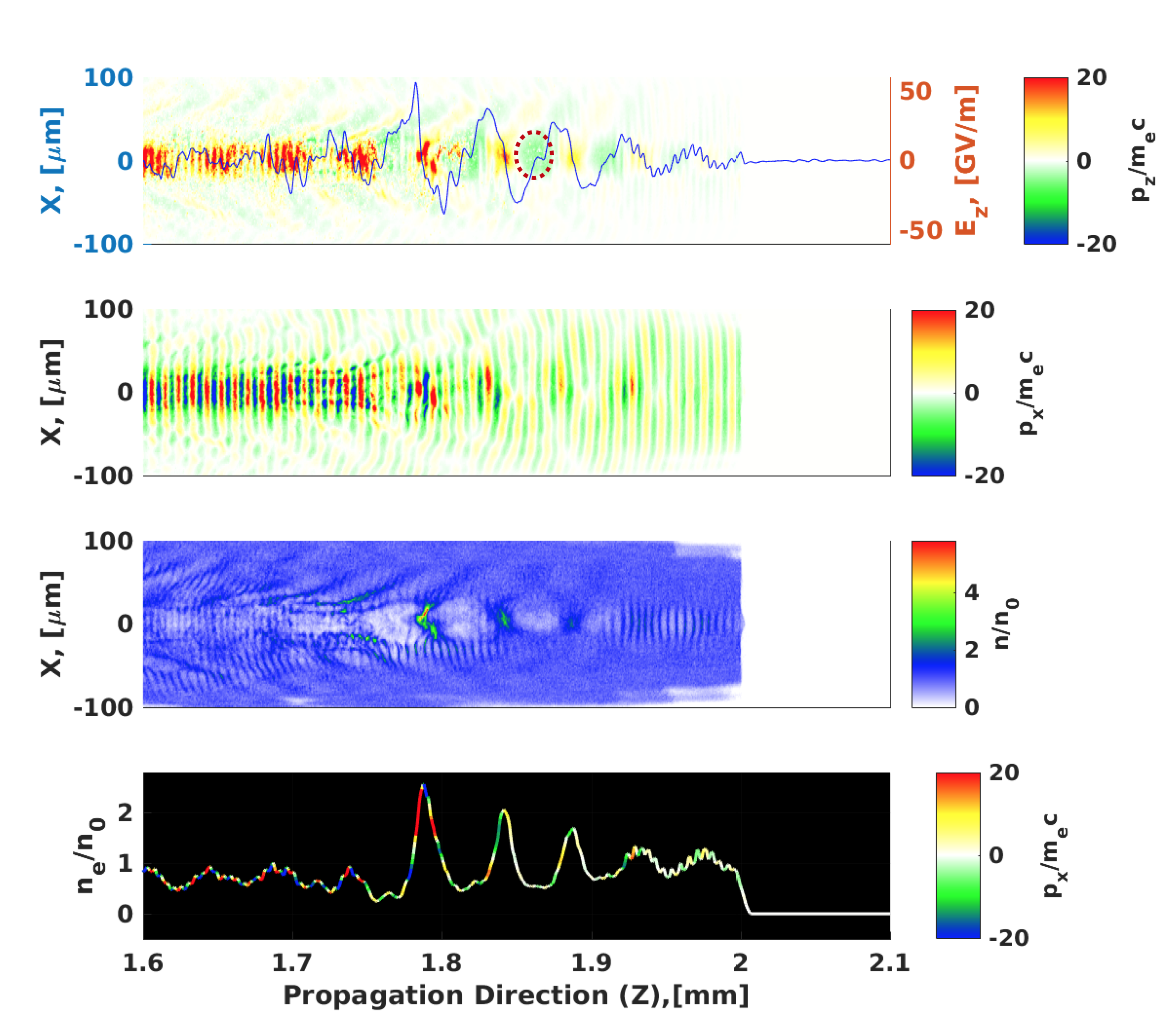}
	\caption{  \raggedright Onset of self-injection. First row : longitudinal momentum distribution is shown on the color plot. Blue curve shows the on-axis wakefield. Secod row : Transverse momentum distribution. Third row : Plasma density distribution. Bottom : Axial line ot of plasma density with the color showing the corresponding value of longitudinal momentum in normalized units. Initial plasma density $(n_0) = 5.0\times 10^{17}$ $cm^{-3}$. Self-injected electrons, having much higher longitudinal momentum compared to background plasma, can be seen before wave-breaking occurs around 1.8 $mm$ (shown in red dotted circle).}
	\label{5e17inj}
\end{figure*}  
\begin{figure}
	\centering
	\includegraphics[width=\linewidth]{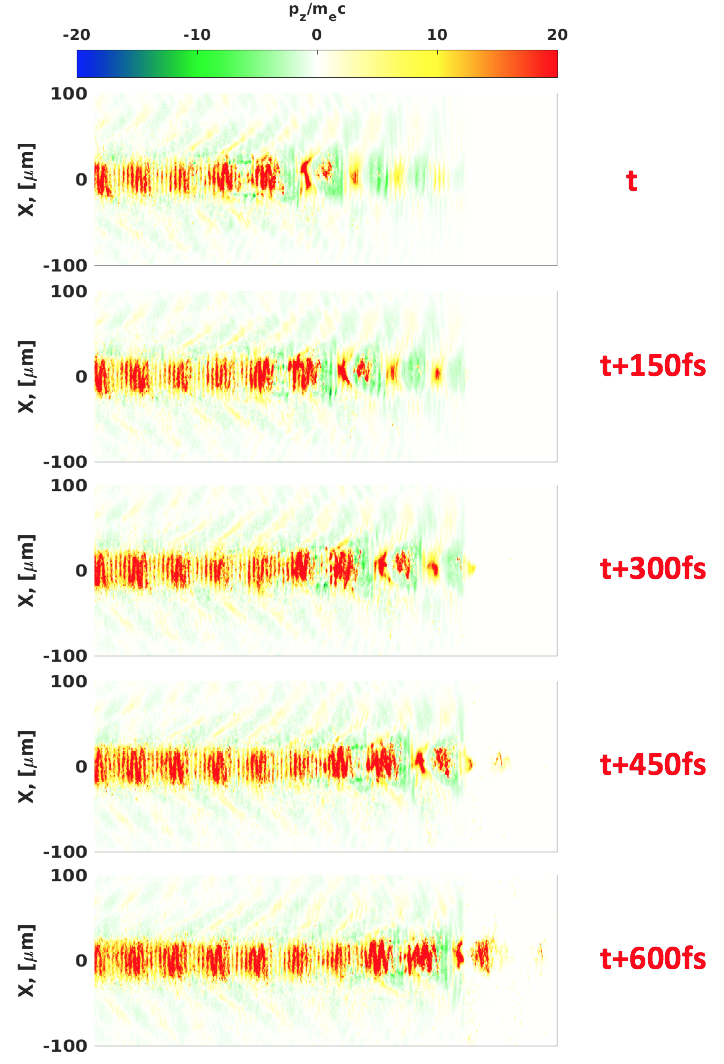}
	\caption{  \raggedright Acceleration of self-injected electrons : Five temporal snapshots of trapped electrons are shown. Self injected electrons propagate with the wake. Plasma ends at 2.0 $mm$. First two injected bunches are shown being ejected out of the plasma region in the bottom plot. Plasma density $(n_e) = 5\times 10^{17}$ $cm^{-3}$.
	}
	\label{inj_acc}	
\end{figure}     
\begin{figure}
	\centering
	\begin{subfigure}{\linewidth}
		\centering
		\includegraphics[width=\linewidth]{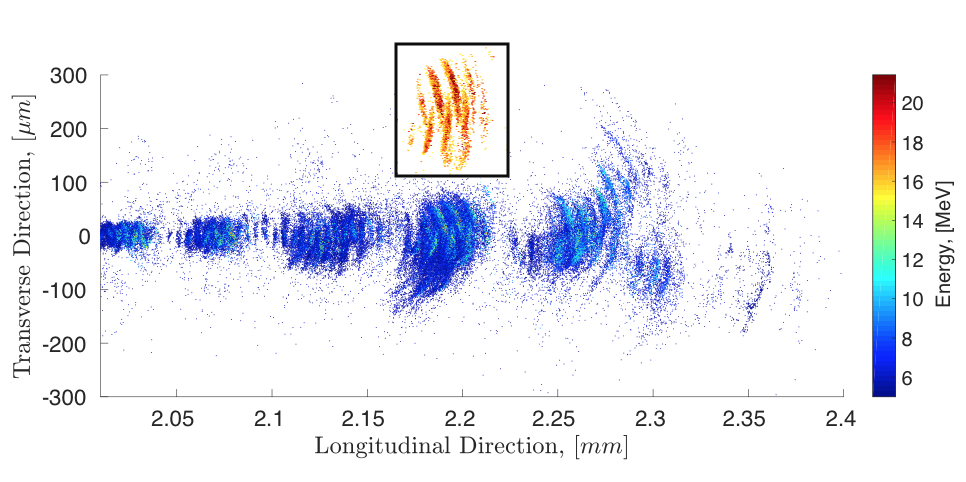}
		\caption{  \raggedright Distribution of accelerated electrons after exiting the plasma}
		\label{5e17DLA}
	\end{subfigure}			
	\begin{subfigure}{0.49\linewidth}
		\centering
		\includegraphics[width=\linewidth]{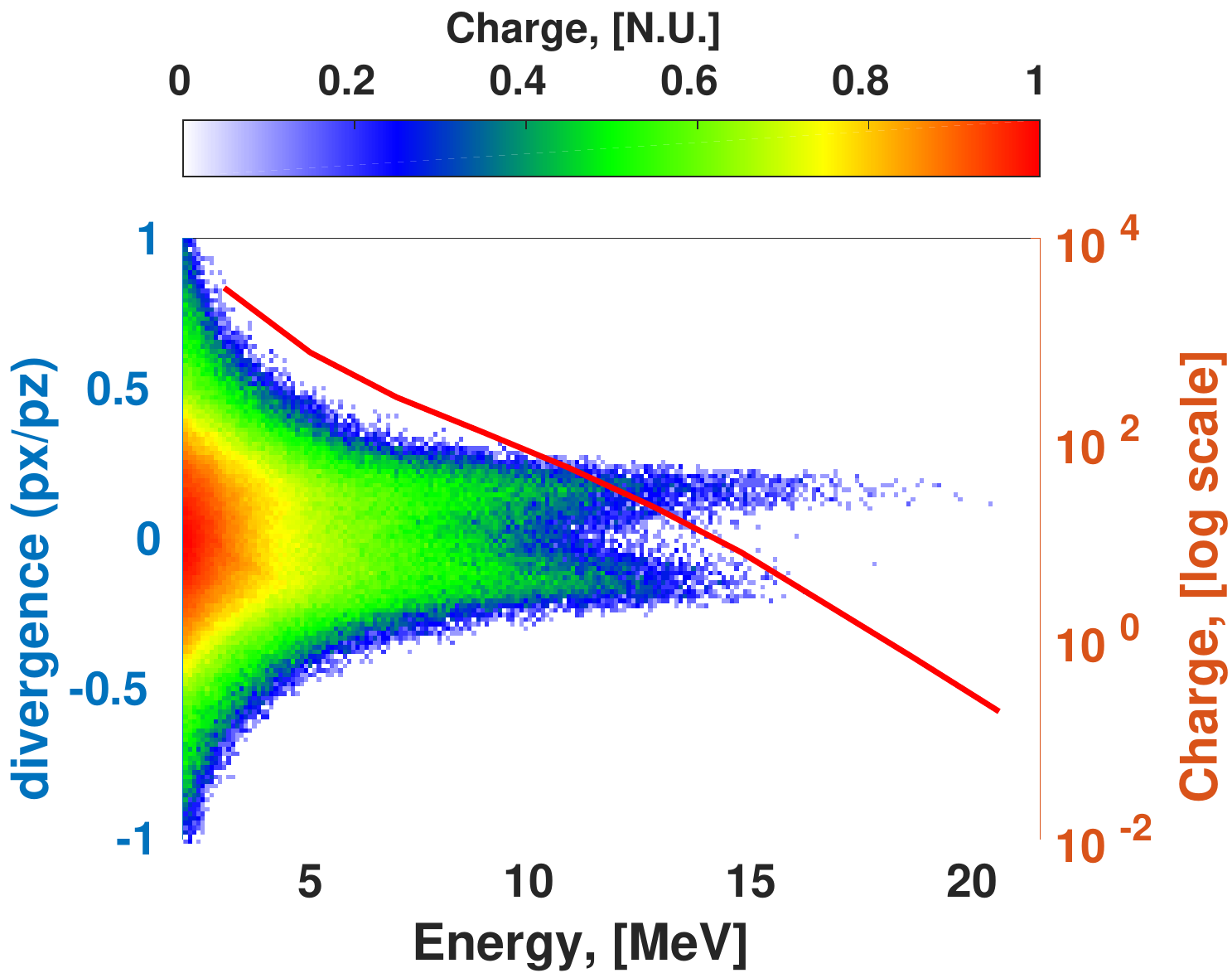}
		\caption{  \raggedright Divergence $(px/pz)$}
		\label{5e17pxpz}
	\end{subfigure}	
	\begin{subfigure}{0.49\linewidth}
		\centering
		\includegraphics[width=\linewidth]{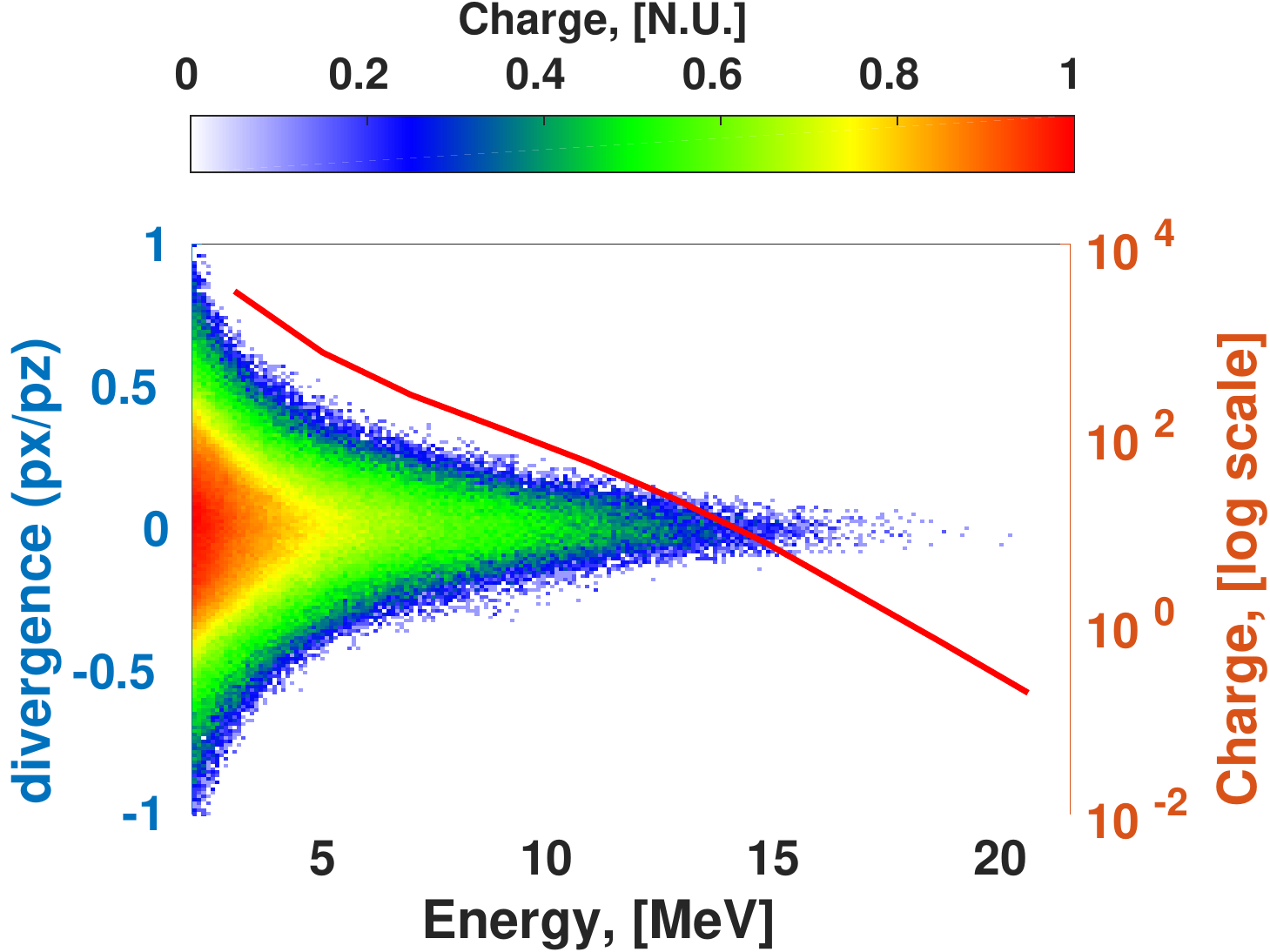}
		\caption{  \raggedright Divergence $(py/pz)$}
		\label{5e17pypz}
	\end{subfigure}	
	\caption{  \raggedright Distribution of accelerated electrons after exiting the plasma. (a) Phase space distribution of first five bunches after being ejected out of plasma. Both DLA and wakefield contribute to acceleration. Highest energy electrons in one of the bunches are filtered out (black box) showing oscillations at laser wavelength. (b) and (c) show angular distribution and energy spectra of these electrons. Few electrons are accelerated to ~22 MeV energy. Divergence (px/pz) shows a fork like structure, a signature of DLA. Plasma density $(n_e) = 5\times 10^{17}$ $cm^{-3}$.     
	}
	\label{5e17DL}	
\end{figure}  
\subsection{Self-injection of electrons and acceleration}
Plasma waves generated in SM-LWFA regime are relativistic, and can reach wave-breaking limits, trap relativistic background electrons, and accelerate them to high energies. Wave-breaking has been suggested to be the mechanism for self-trapping to occur. For cold plasma in 1D, the wave-breaking limit, $E_{wb}$ is expressed in equation \ref{wb}.
\begin{equation}
E_{wb} = \sqrt{2(\gamma_p-1)}E_0.
\label{wb}
\end{equation}
Here 
\begin{equation}
\gamma_p = 1/\sqrt{1-(v_p/c)^2}
\label{gamma_p}
\end{equation}
is the Lorentz factor associated with the relativistic plasma wave and $E_0 = m_ec\omega_p/e$ is the accelerating field associate with linear plasma wave with phase velocity $v_p$. The wave breaking limit for warm plasma has been approximated using 1D fluid theory and is given by:
\begin{equation}
E_{wb} = (2\gamma_\perp(\gamma_p-1) - \beta_p^2\gamma_p\gamma_\perp(8\delta_{th}/3-2\delta_{th}^2))^{1/2}E_0.
\label{wbth}
\end{equation}
Here $\delta_{th} = (3\beta_{th}^2\gamma_p^2/\gamma_\perp^2\beta_p^2)^{1/4}$, $\gamma_\perp = 1 + a_0^2$, $c\beta_{th} = (k_BT_0/m)^{1/2}$ where $k_B$ is the Boltzmann constant and $T_0$ is the initial electron plasma temperature. Thermal and 3D effects are known to reduce the threshold for self-injection and it has been shown to occur at wakefield amplitudes 10 - 30$\%$ of wave breaking limit in self-modulated regime experiments\cite{Esarey1998}. The process of self-injection in our simulation is demonstrated in figure \ref{5e17inj}. Plasma density distribution plot (third panel from top) and its axial line out (line plot in the bottom pannel) show a few plasma wave periods in the front before the wave breaks. Longitudinal momentum distribution is shown in the top row and its axial distribution is color mapped on the axial plasma density distribution in the bottom row. Self-injected electrons having a large positive longitudinal momentum (red dotted circle) can be seen to appear before wave-breaking begins around $1.8$ $mm$. For a plasma density of $5.0\times 10^{17}$ $cm^{-3}$, the wave breaking limit calculated using equations \ref{wb} and \ref{wbth} are 195 and 169 $GV/m$ respectively. The amplitude of local longitudinal field $(E_z) = 40$ $GV/m$ in our simulations is $ \sim20\%$ of the  cold wave breaking limit and $\sim 24\%$ of the warm wave breaking limit. 
  
Acceleration of trapped electrons are demonstrated in figure \ref{inj_acc}. Five temporal snapshots separated by $150$ $fs$ are shown. A bunch trapped in the accelerating phase of the wake has a large positive longitudinal momentum (first row of figure \ref{inj_acc}). As the wakes propagate to the right, trapped electrons move forward with the wake. Final row shows first two trapped bunches ejected out of the plasma.  

The longitudinal electric field created by the self-modulated wakes can accelerate electrons to relativistic energies. In addition, if the pulse duration of the drive laser is long enough to overlap the trapped electrons, another acceleration mechanism known as direct laser acceleration (DLA) can be induced\cite{ShawDLA,ShawDLA1,Shaw_2016}. 

In DLA, the electrons which are overlapped by the laser pulse, gain transverse momentum from the laser pulse and undergo betatron oscillations. When betatron frequency, $\omega_\beta = \omega_p/\sqrt{2\gamma}$, matches the laser frequency witnessed by the electrons, these electrons gain longitudinal momentum through $v\times B$ force. The nature and trajectory of the accelerated electrons is determined by the combination of these two accelerating mechanisms. 

Figure \ref{5e17DL} shows distribution and angular spread of accelerated bunches ejected out of the plasma region. These bunches are separated by plasma period and have energies reaching up to 22 $MeV$. The highest energy electrons, when filtered out from one of the bunches (shown in the black box in figure \ref{5e17DLA}), show structure resulting from oscillations at laser wavelength. Figures \ref{5e17pxpz} and \ref{5e17pypz} show the angular spread. Highest energy electrons are closer to the center, as they experience minimum energy spread. While undergoing betatron oscillations, these electrons spend maximum time at the extrema resulting in a fork-like structure as shown in figure \ref{5e17pxpz}.

\begin{figure*}
	\centering
	\begin{subfigure}{0.49\linewidth}
		\centering
		\includegraphics[width=\linewidth]{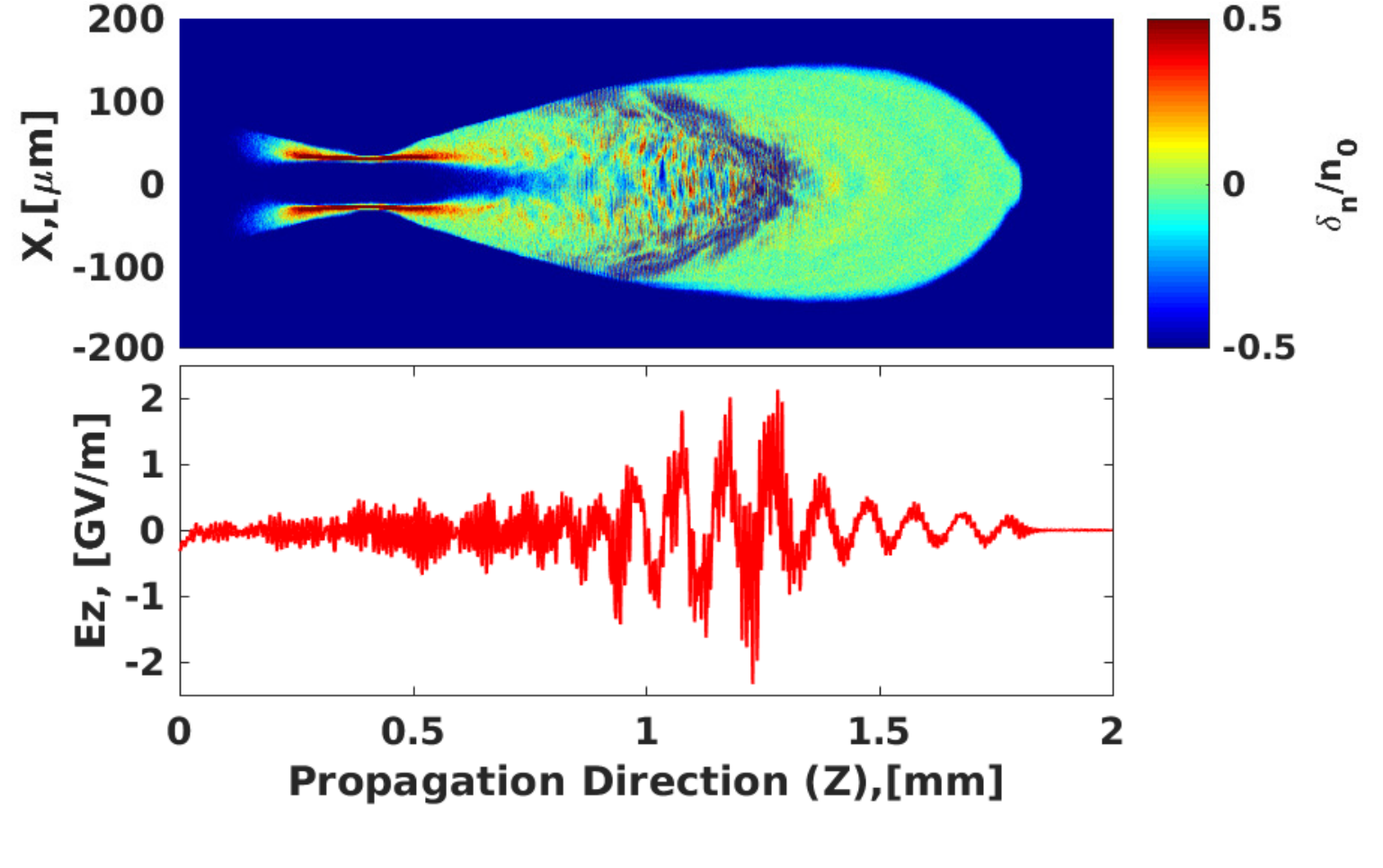}
		\caption{  \raggedright $n_e = 8\times 10^{16}$ $cm^{-3}$}
		\label{8e16}
	\end{subfigure}
	\begin{subfigure}{0.49\linewidth}
		\centering
		\includegraphics[width=\linewidth]{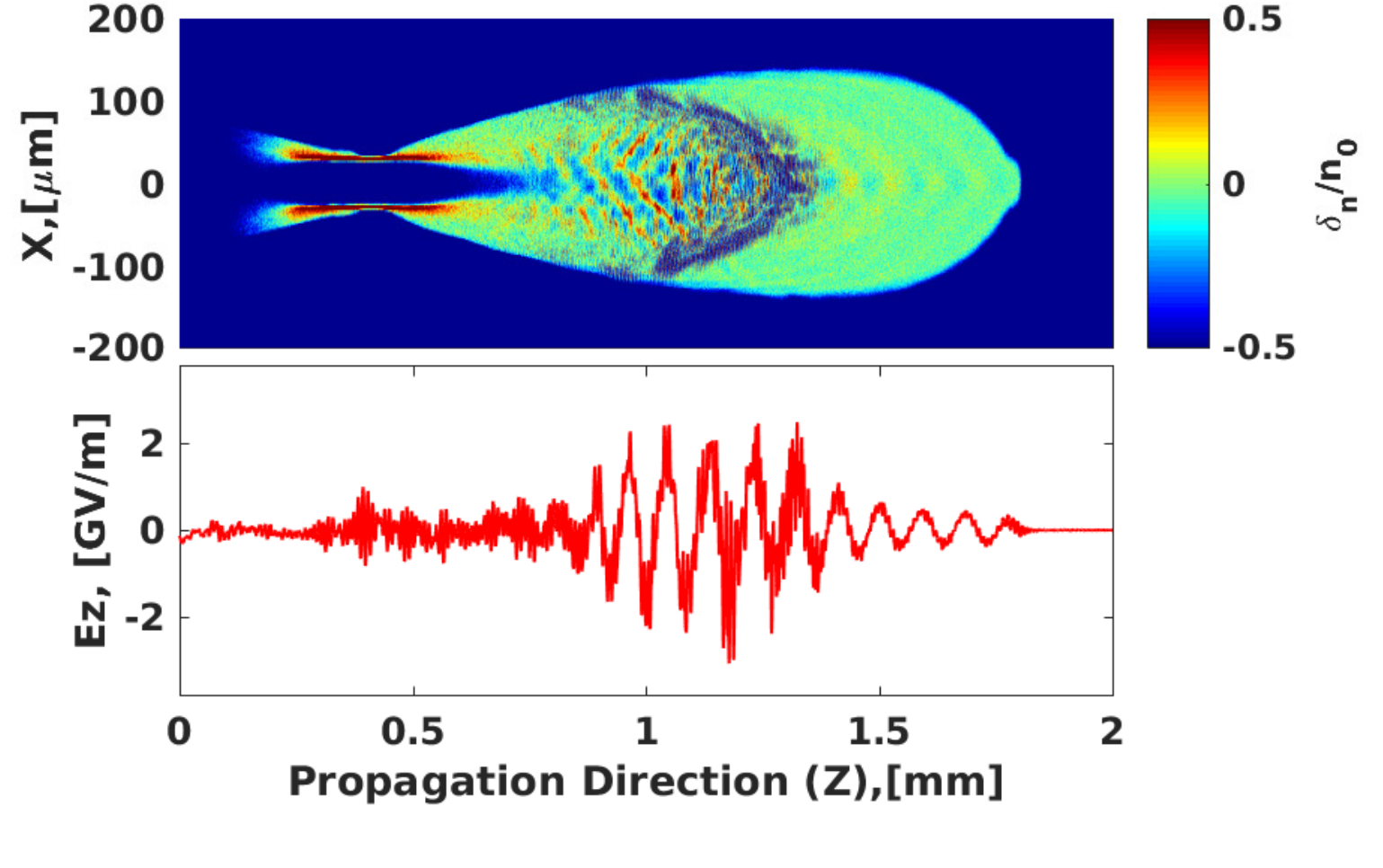}
		\caption{  \raggedright $n_e = 1\times 10^{17}$ $cm^{-3}$}
		\label{1e17}
	\end{subfigure}			
	\begin{subfigure}{0.49\linewidth}
		\centering
		\includegraphics[width=\linewidth]{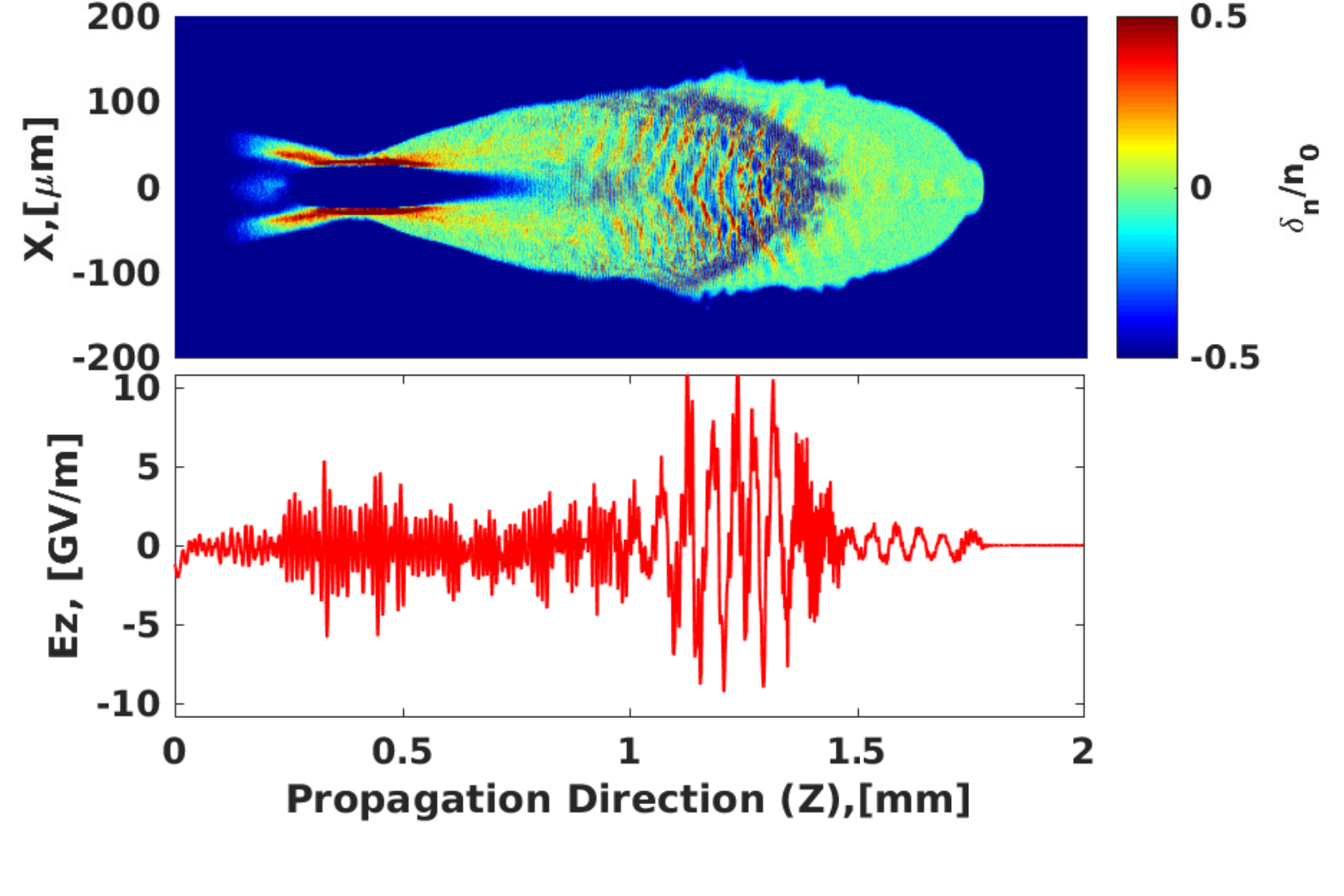}
		\caption{  \raggedright $n_e = 3\times 10^{17}$ $cm^{-3}$}
		\label{3e17ch}
	\end{subfigure}
	\begin{subfigure}{0.49\linewidth}
		\centering
		\includegraphics[width=\linewidth]{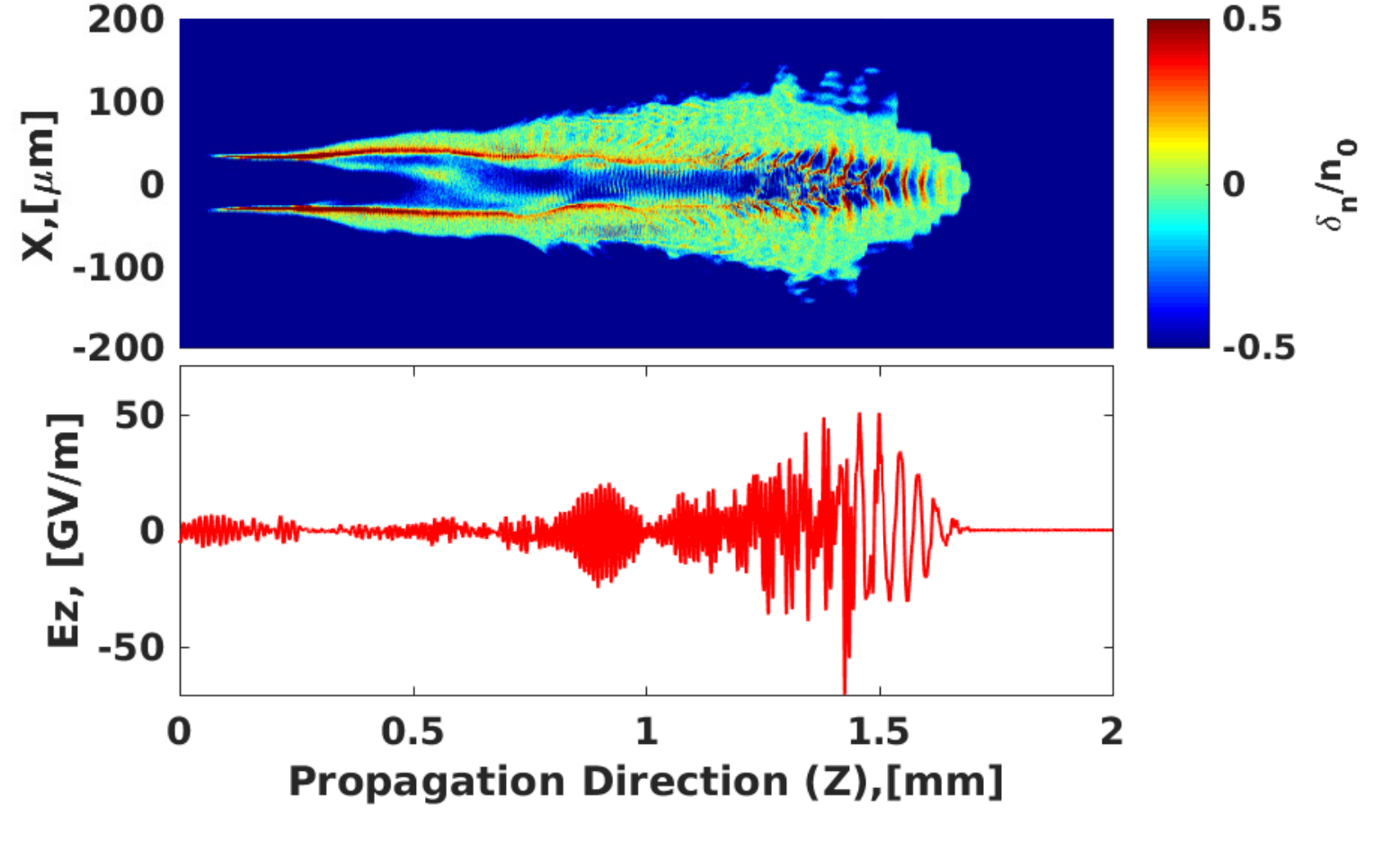}
		\caption{  \raggedright $n_e = 7.5\times 10^{17}$ $cm^{-3}$}
		\label{7.5e17ch}
	\end{subfigure}			
	\caption{  \raggedright Effect of variation in plasma density : Structure of the self-modulated wakes and corresponding longitudinal field is shown for five different plasma densities. Linearly polarized CO$_2$ laser is propagating from left to right. $X$ is the direction of linear polarization of the laser pulse. Ionization model is used and ions are assumed to be mobile. Wakefield amplitude increases with increase in plasma density. High wakefield generation at higher densities leads to transition into wave-breaking sooner for higher density cases.
	}
	\label{self_inj_wake}
	
\end{figure*}  
\begin{figure*}
	\centering
	\begin{subfigure}{0.4\linewidth}
		\centering		
		\includegraphics[width=\linewidth]{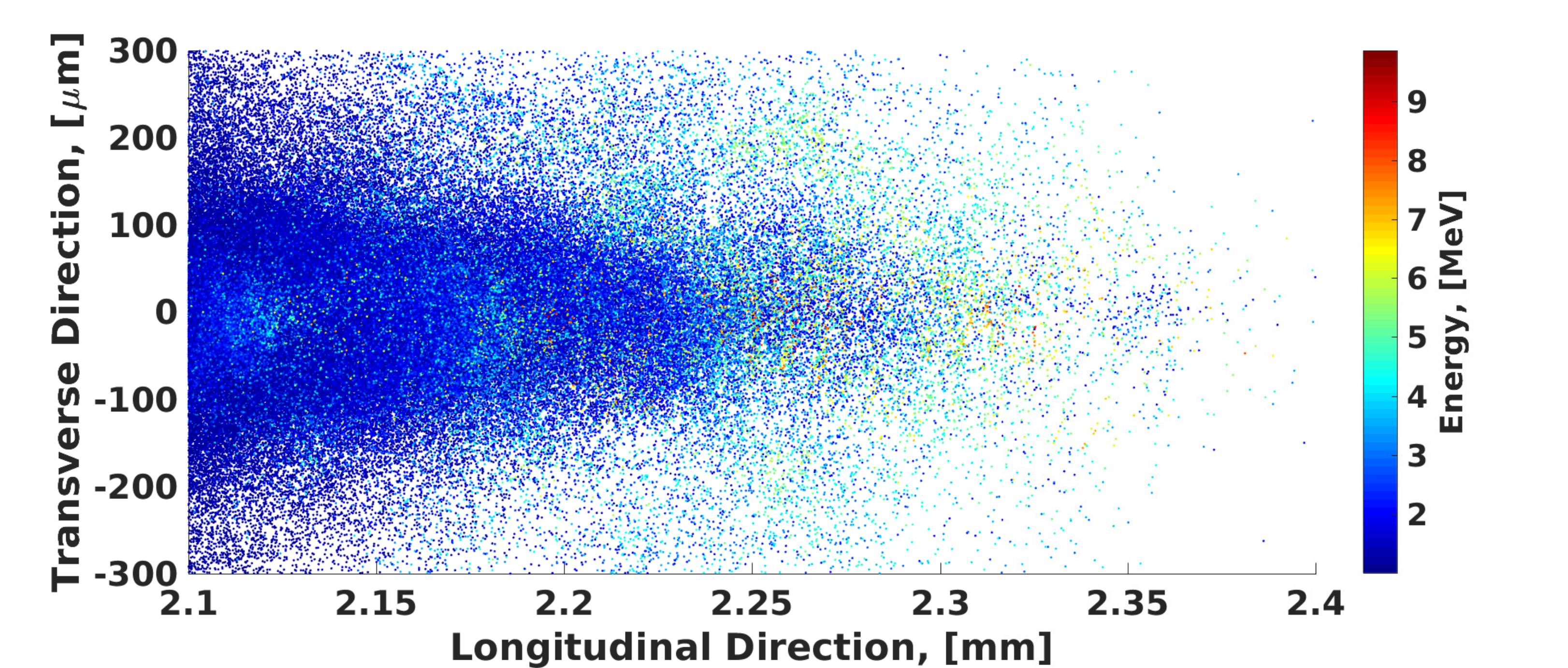}
		\caption{  \raggedright $n_e = 3\times 10^{17}$ $cm^{-3}$}
		\label{3e17acc}
	\end{subfigure}			
	\begin{subfigure}{0.25\linewidth}
		\centering
		\includegraphics[width=\linewidth]{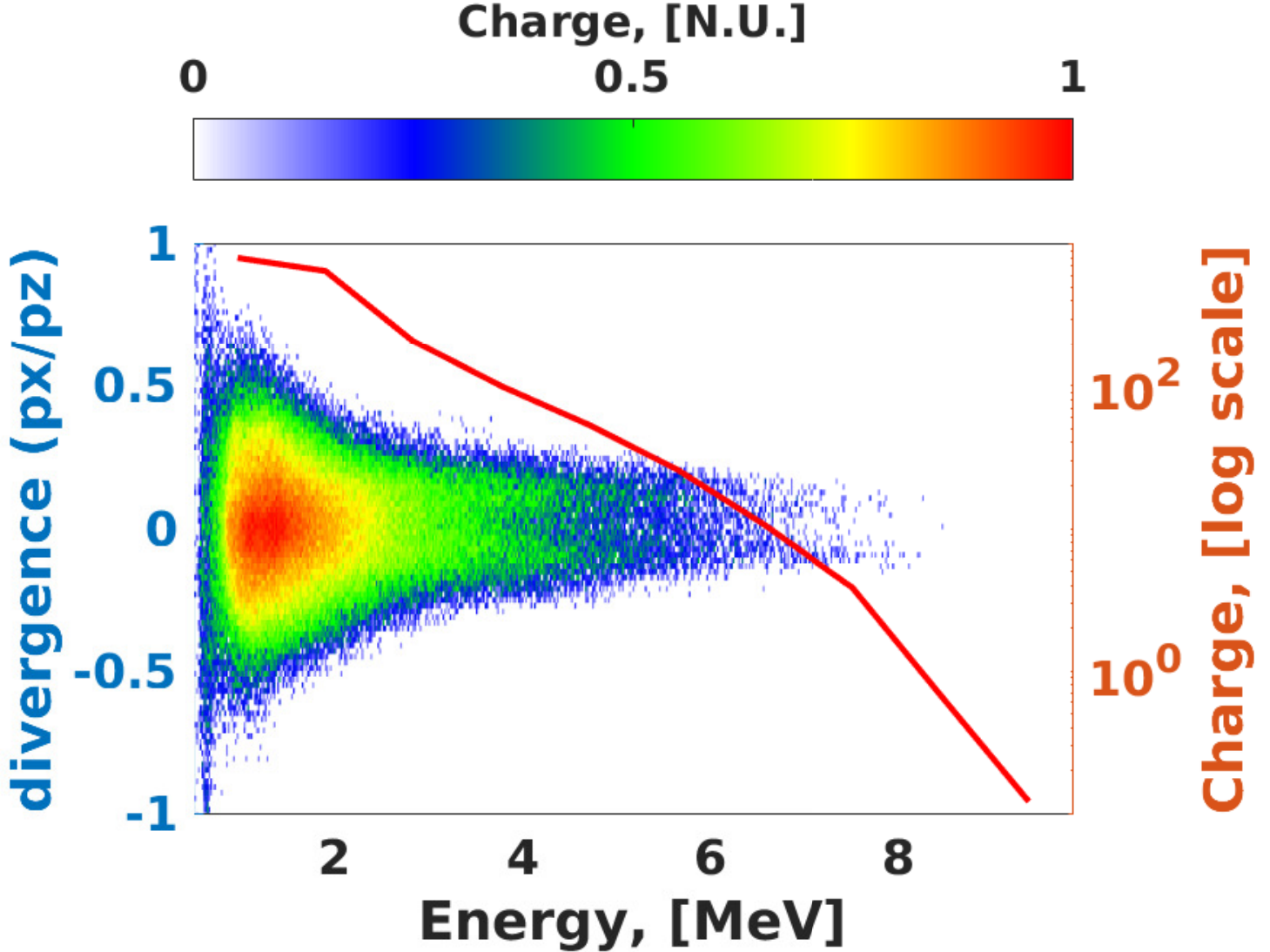}
		\caption{  \raggedright $n_e = 3\times 10^{17}$ $cm^{-3}$}
		\label{3e17pxpz}
	\end{subfigure}	
	\begin{subfigure}{0.25\linewidth}
		\centering
		\includegraphics[width=\linewidth]{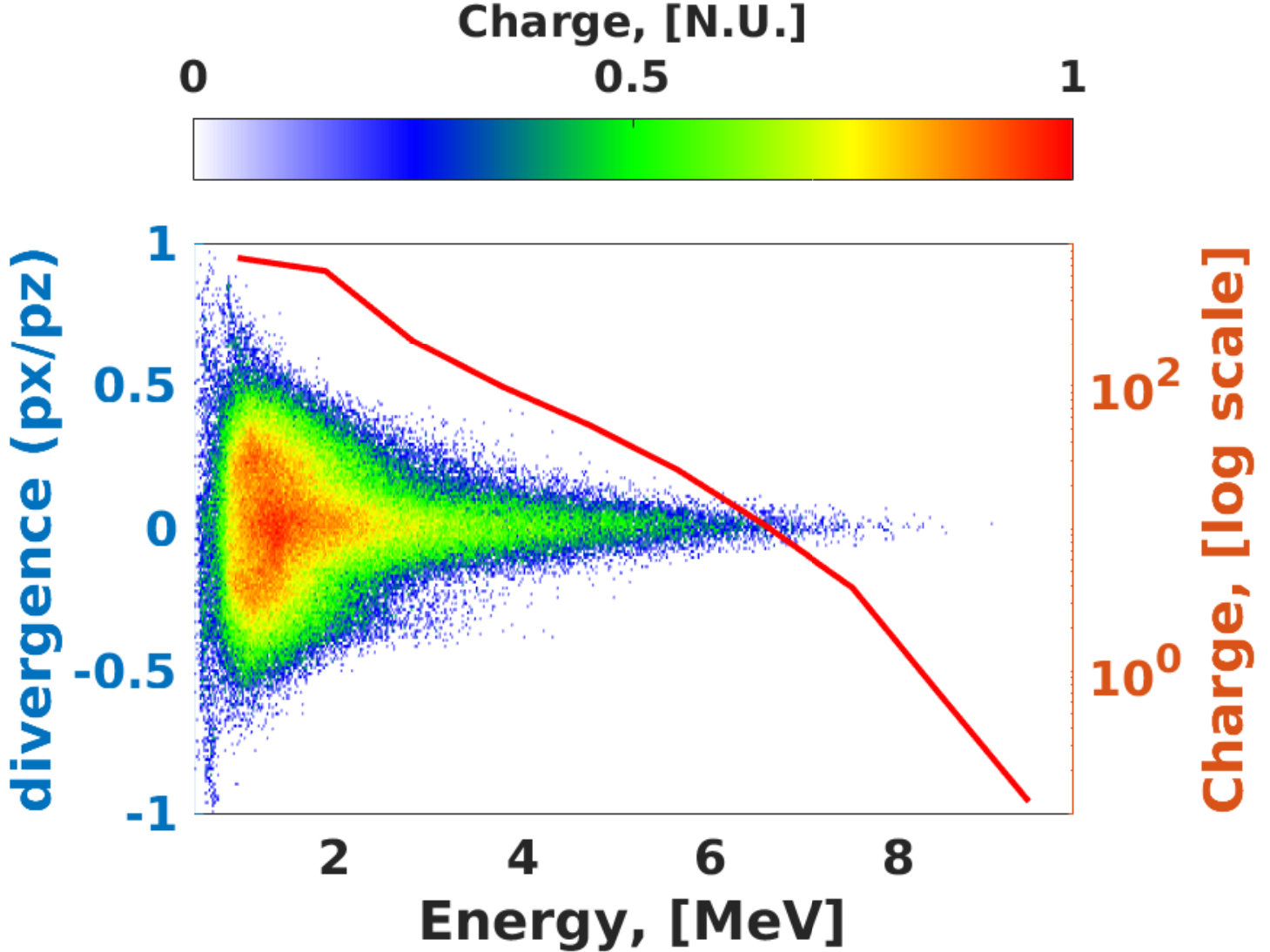}
		\caption{  \raggedright $n_e = 3\times 10^{17}$ $cm^{-3}$}
		\label{3e17pypz}
	\end{subfigure}	
	\begin{subfigure}{0.4\linewidth}
		\centering		
		\includegraphics[width=\linewidth]{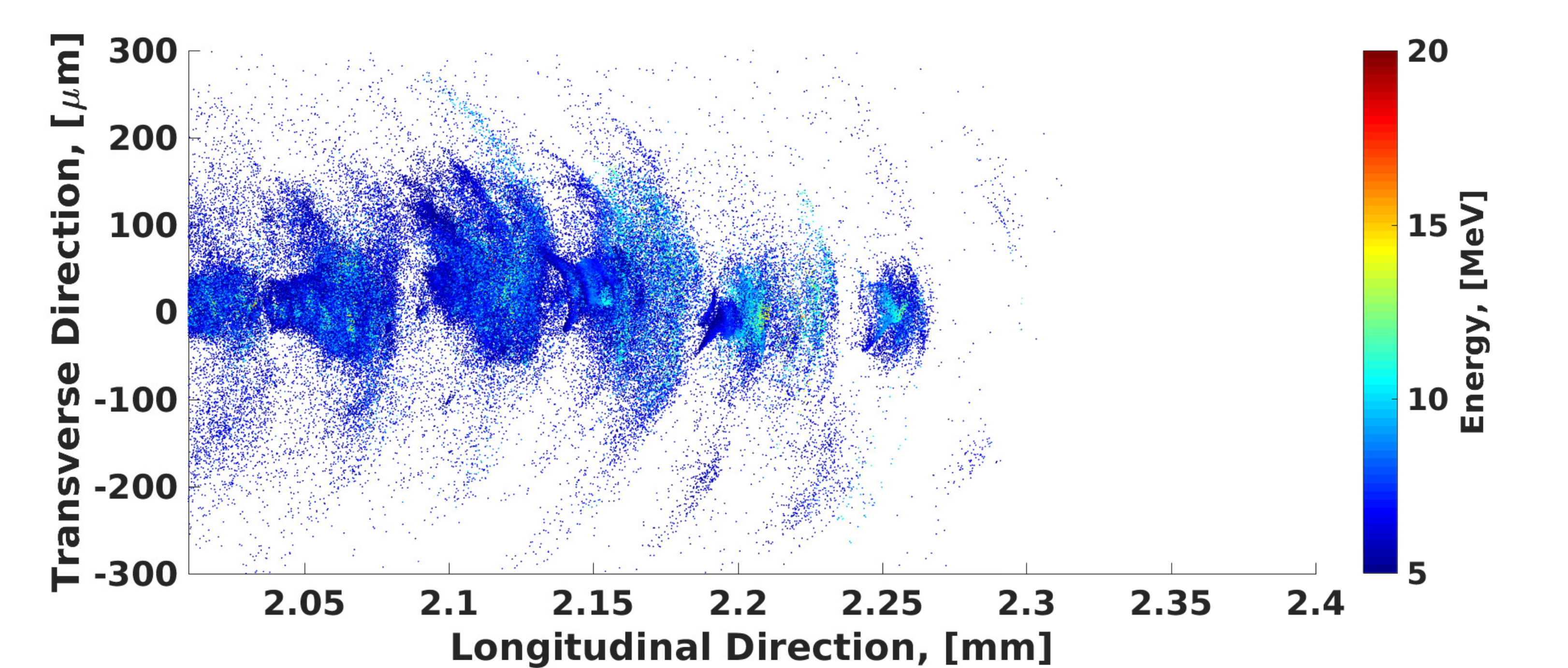}
		\caption{  \raggedright $n_e = 7.5\times 10^{17}$ $cm^{-3}$}
		\label{7.5e17acc}
	\end{subfigure}			
	\begin{subfigure}{0.25\linewidth}
		\centering
		\includegraphics[width=\linewidth]{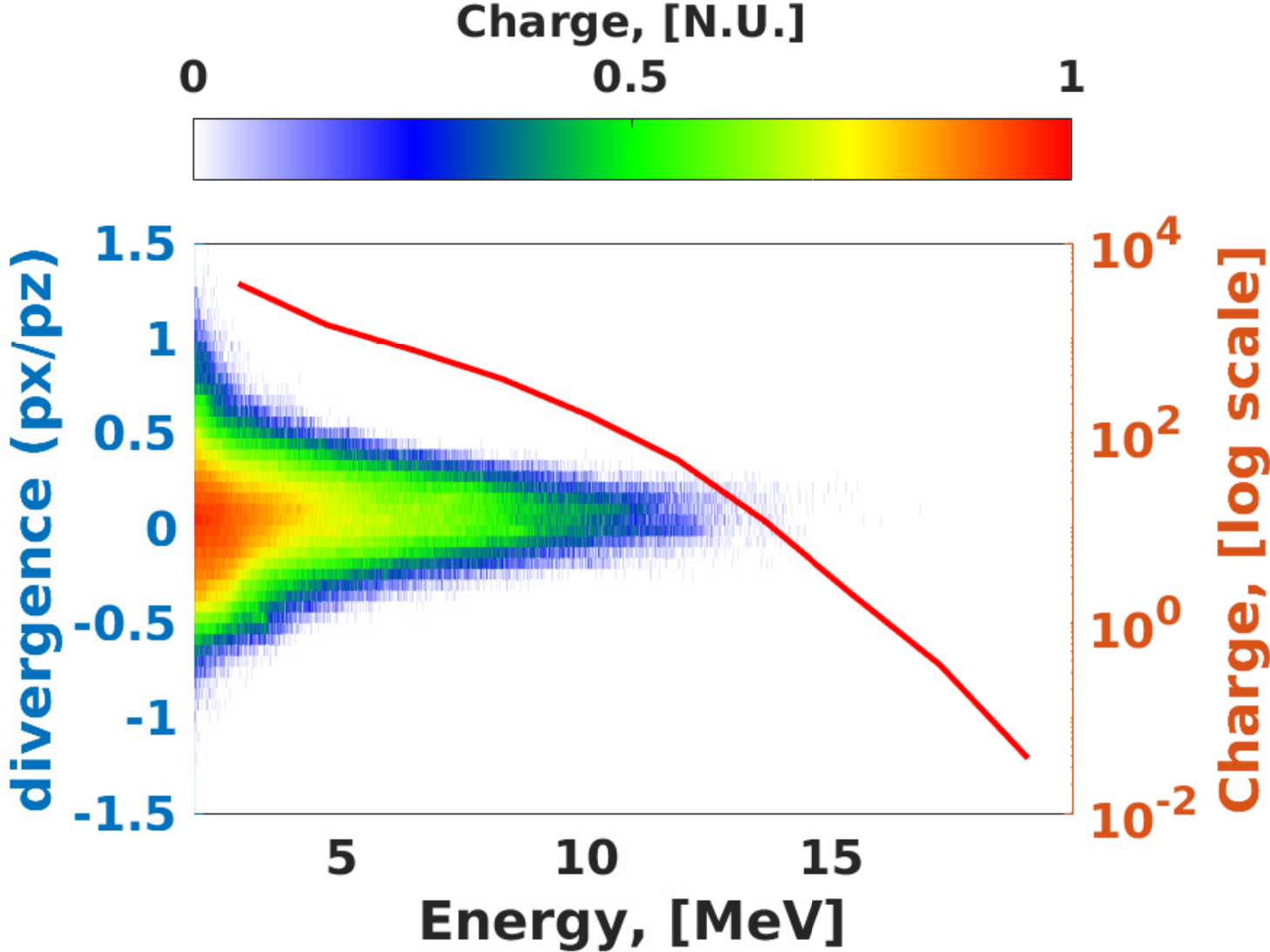}
		\caption{  \raggedright $n_e = 7.5\times 10^{17}$ $cm^{-3}$}
		\label{7.5e17pxpz}
	\end{subfigure}	
	\begin{subfigure}{0.25\linewidth}
		\centering
		\includegraphics[width=\linewidth]{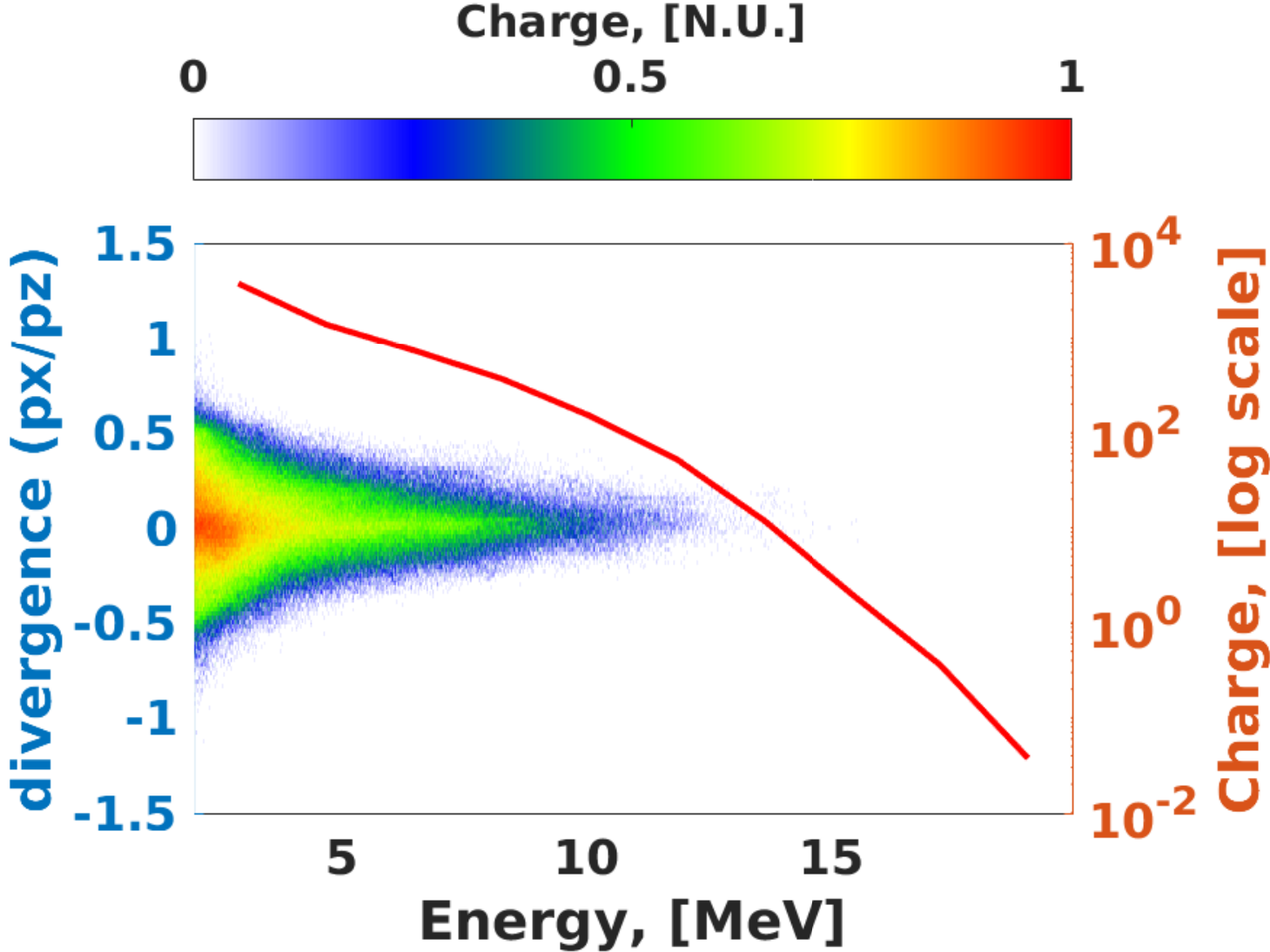}
		\caption{  \raggedright $n_e = 7.5\times 10^{17}$ $cm^{-3}$}
		\label{7.5e17pypz}
	\end{subfigure}	
	\caption{  \raggedright Distribution of accelerated electrons after exiting the plasma with different densities. (a) Phase space distribution of accelerated electrons exited from plasma. Both DLA and wakefield contribute to acceleration. (b), (c), (e) and (f) show angular distribution and energy spectra of these electrons. Divergence (px/pz) shows a clear fork like structure for $n_e = 3\times 10^{17}$ $cm^{-3}$, however for $n_e = 7.5\times 10^{17}$ $cm^{-3}$, the fork structure is suppressed as the wakefield contribution increases. Up to 20 $MeV$ electrons are seen in the highest density case.     
	}
	\label{eject}	
\end{figure*} 

\subsection{Effect of variations in plasma density}  
Here we present the result of variation in plasma density on the nature of self-injected electrons for a fixed laser energy in the LWIR regime. We also determine the threshold plasma density for self-injection for the given laser energy. We start with a hydrogen plasma with peak density of $7.5 \times 10^{17}$ electrons $cm^{-3}$ and gradually reduce the density to $n_e = 8 \times 10^{16}$ $cm^{-3}$ while keeping all other physical and numerical parameters constant.

Figure \ref{self_inj_wake} shows the structure of the self-modulated wakes and corresponding longitudinal field for plasma densities $8.0 \times 10^{16}$, $1.0 \times 10^{17}$, $3.0 \times 10^{17}$, and $7.5 \times 10^{17}$ $cm^{-3}$. As the plasma density increases, an increase in corresponding wakefield generated is seen as expected. Moreover, the peak accelerating field is reached sooner in the interaction. This has been attributed to stronger relativistic self-focusing in higher density cases. The refractive index of plasma is given by:
\begin{equation}
\eta_R = \sqrt{1-K\frac{\lambda_{CO\textsubscript{2}}n_e}{\sqrt{1+\frac{a_0^2}{2}}}},
\end{equation}
where $K$ is a constant, $\lambda_{CO\textsubscript{2}}$ is the laser wavelength, $n_e$ is the plasma electron density and $a_0$ is the normalized vector potential of the laser.
Since, for a given laser $a_0$, a higher $n_e$ causes the laser to focus more, resulting in a stronger wakefield generation. In the highest density case, $n_e = 7.5\times 10^{17}$ $cm^{-3}$, peak accelerating field reaches $74$ $GV/m$. This triggers the process of self-injection of background electrons sooner. 

Phase space distribution of electrons ejected out of the plasma range is shown in figure \ref{eject}. For higher density cases, accelerated electrons reach up to $\sim 22$ $MeV$ of energy and are bunched together, separated by plasma wavelengths. In the low density case, wakefield amplitude is smaller and electrons with high angular spread and low energy are generated. For $n_e = 1\times 10^{17}$ $cm^{-3}$, very few scattered accelerated electrons were observed, and for $n_e = 8\times 10^{16}$ $cm^{-3}$ self-injection does not occur.

Figures \ref{3e17pxpz}, \ref{3e17pypz}, \ref{7.5e17pxpz}, and \ref{7.5e17pypz} show the angular spread of the accelerated electrons. Divergence angle is calculated as the ratio $(v_x/v_{||})$ and  $(v_y/v_{||})$. Here $x$ is the direction of linear polarization of the laser pulse and $y$ is the other transverse coordinate. High energy electrons can be seen close to the center, whereas low energy electrons have a much larger angular spread. Divergence (px/pz) shows a clear fork like structure for $n_e = 3\times 10^{17}$ $cm^{-3}$, however for $n_e = 7.5\times 10^{17}$ $cm^{-3}$, the fork structure is suppressed as the wakefield contribution increases.

\subsection{Simulation at higher energy}
\begin{figure*}
	\centering
		\includegraphics[width=\linewidth]{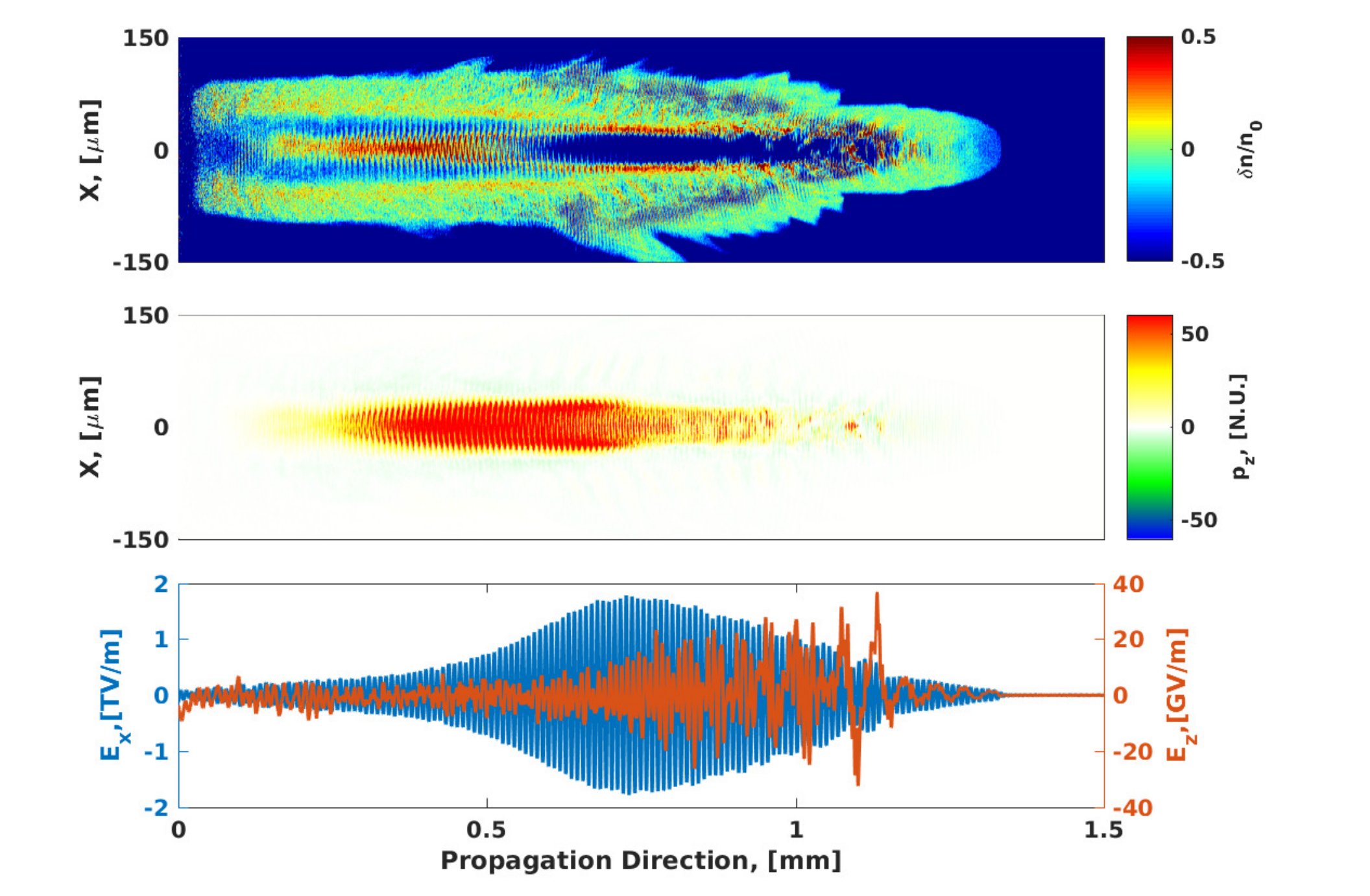}			
	\caption{  \raggedright Simulation with $15 J, 2 ps$ laser : Top panel shows few periods of plasma wakes followed by a self-formed channel. Middle panel shows corresponding longitudinal momentum distribution. On-axis transverse and longitudinal fields are shown in the bottom panel. Electrons trapped in the channel gain forward momentum from the laser pulse and are accelerated to multi-MeV energies. 
	}
\label{15J_chpz}
\end{figure*}
\begin{figure*}
	\centering
	\begin{subfigure}{0.4\linewidth}
		\centering
		\includegraphics[width=\linewidth]{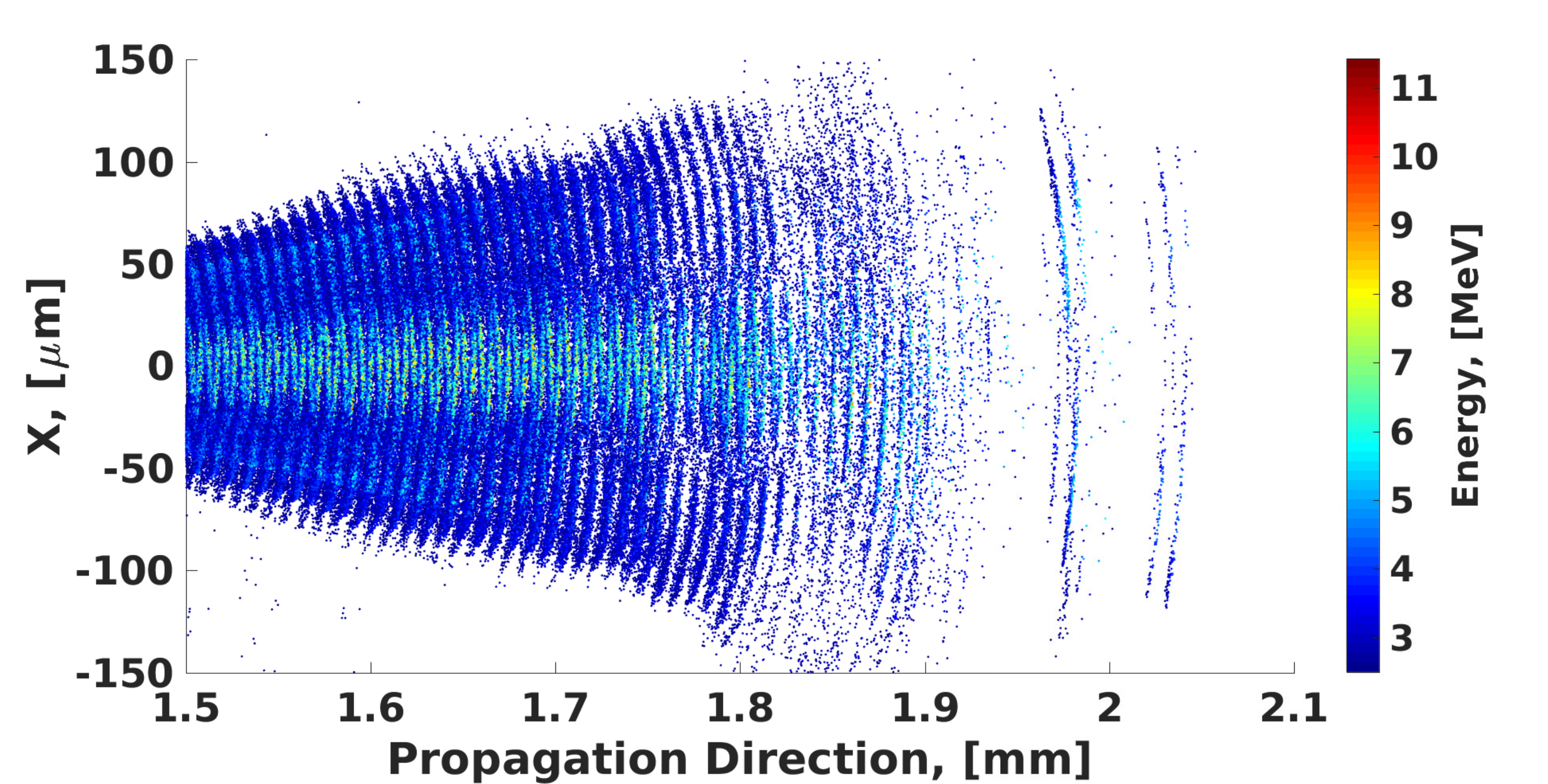}
		\caption{  \raggedright Distribution of accelerated electrons ejected out the plasma}
		\label{15J_en}
	\end{subfigure}	
	\begin{subfigure}{0.29\linewidth}
		\centering
		\includegraphics[width=\linewidth]{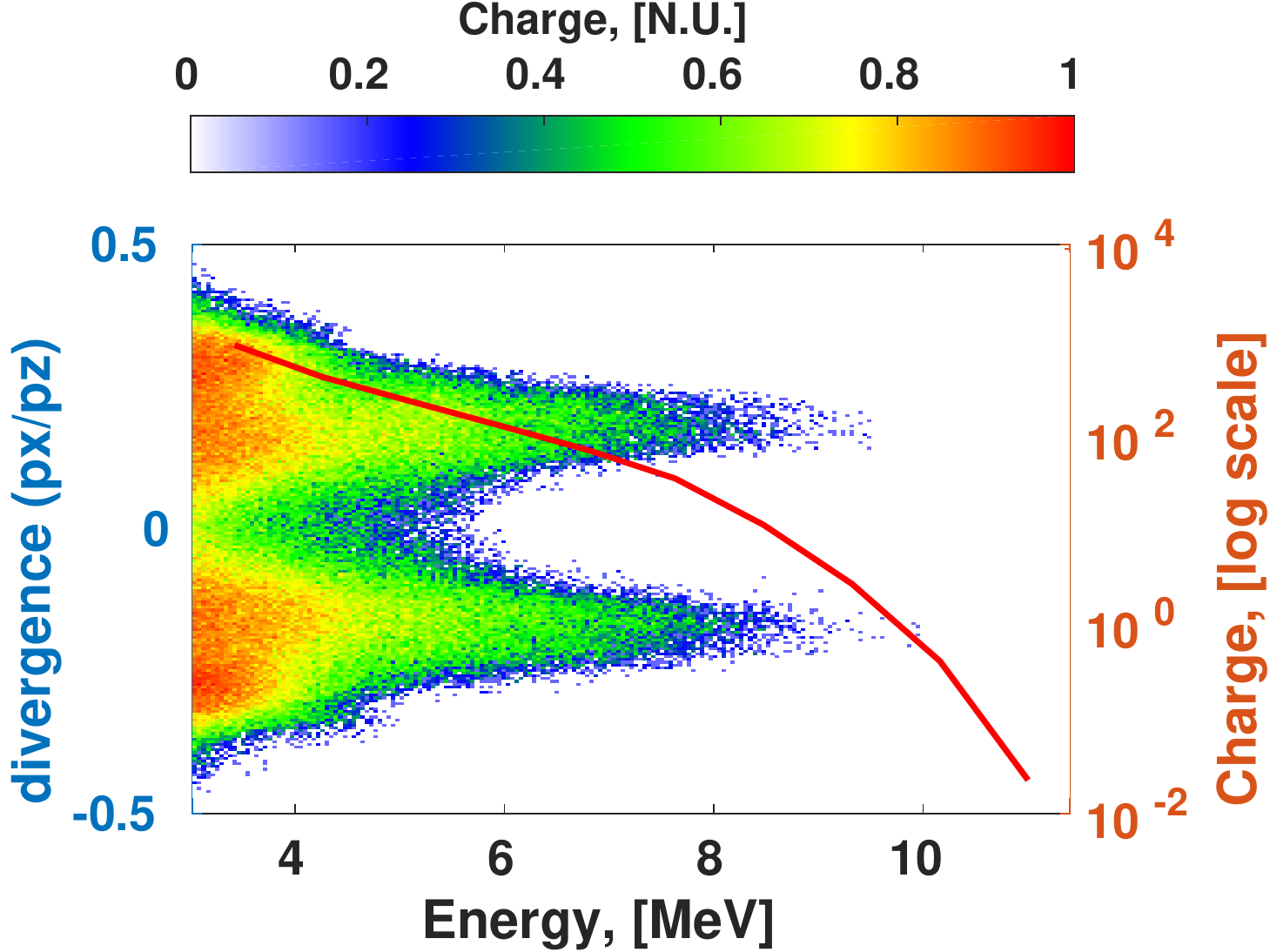}
		\caption{  \raggedright Divergence $(px/pz)$}
		\label{15J_pxpz}
	\end{subfigure}	
	\begin{subfigure}{0.29\linewidth}
		\centering
		\includegraphics[width=\linewidth]{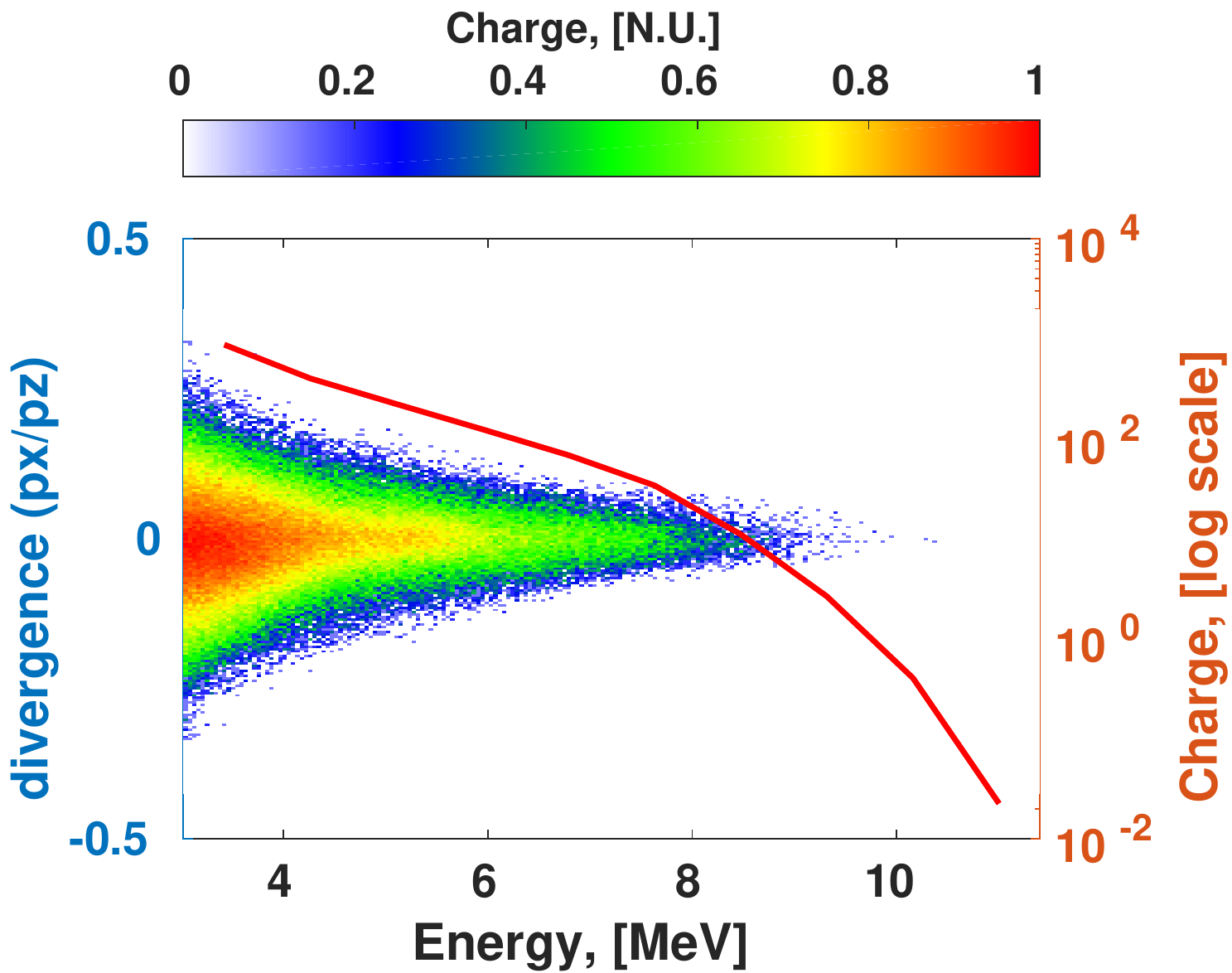}
		\caption{  \raggedright Divergence $(py/pz)$}
		\label{15J_pypz}
	\end{subfigure}		
	\caption{  \raggedright Simulation with 15 J, 2 ps laser: (a) Distribution of accelerated electrons ejected out of the plasma region. (b) The divergence in the plane of laser polarization $(px/pz)$, shows a very prominent fork-like structure. Both divergence plots show that the highest energy electrons come from the central region of the channel where the electrons performing betatron oscillation have the maximum velocity.
	}
	\label{15J_eject}	
\end{figure*}

Now we present the effect of increasing the laser energy. Laser pulse with 15 J of energy at the focus position was injected from the left boundary of the simulation box. Beam waist, $w_0 = 50$ $\mu m$ was chosen to give a moderate $a_0$ value of 3.3. Plasma length of $1.5$ $mm$ with $250$ $\mu m$ ramps on either side was simulated with the peak density of $5.0 \times 10^{17}$ $cm^{-3}$. Figure \ref{15J_chpz} shows the structure of plasma wakes and corresponding distribution of longitudinal momentum. Wave breaking occurs very early in the interaction behind relatively fewer plasma wave periods and triggers the process of self-injection. Laser undergoes relativistic channeling behind the broken wave and the electrons trapped in the channel move forward while undergoing betatron oscillations. Due to resonance between betatron and laser frequencies, there is an efficient transfer of energy from laser to the trapped electrons. Energy gain from self-modulated wakefield seem to play only a minor role. Since the wave breaking and hence onset of self-channeling is reached sooner in this case compared to simulations at lower energies, the length of the channel is maximized and laser overlaps a larger portion of trapped electrons. 
 
Figure \ref{15J_eject} shows the distribution of accelerated electrons ejected out of the plasma region. The two rightmost bunches emerge from the two self-modulated wake buckets and are followed by a stream of electrons emerging from the channel. These electrons gain enegy from DLA process and get ejected depicting structures oscillating at laser wavelength as shown in figure \ref{15J_en}. The divergence in the plane of laser polarization $(px/pz)$, shows a very prominent fork-like structure. Both divergence plots show that the highest energy electrons come from the central region of the channel where the electrons performing betatron oscillation have the highest velocity. The highest energy reached by the electrons in this case is $\sim 12$ $MeV$, lower than the corresponding simulation with a $4$ J, $2$ ps laser pulse. Even though the laser energy is nearly four times higher, the accelerating field generated in this case lower (Figure \ref{15J_chpz} bottom panel) than the $4$ J, $2$ ps simulation.

\begin{figure*}
	\centering
	\begin{subfigure}{0.25\linewidth}
	\includegraphics[width=4.5cm, height=5cm]{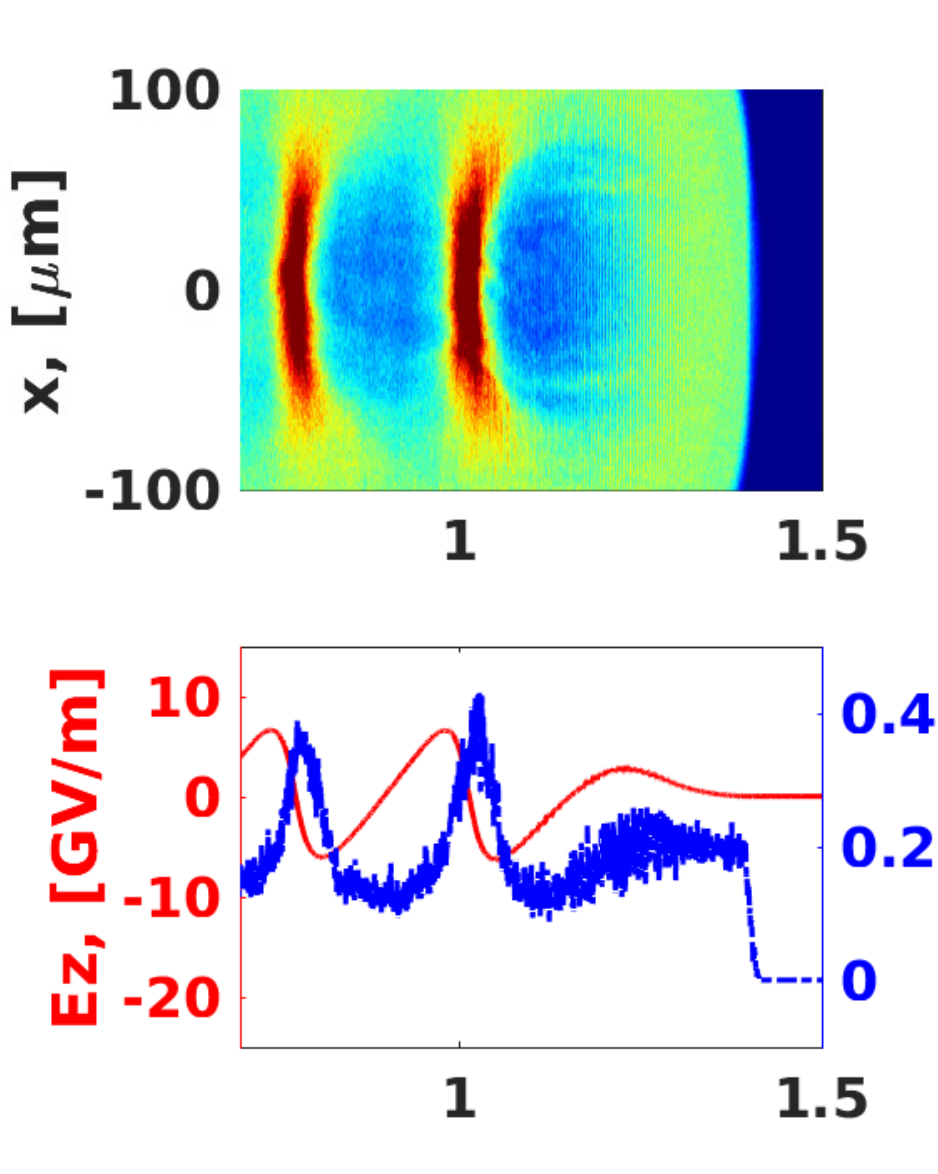}
	\caption{ }
	\label{a}
    \end{subfigure}	
    \begin{subfigure}{0.24\linewidth}
    	\includegraphics[width=\linewidth]{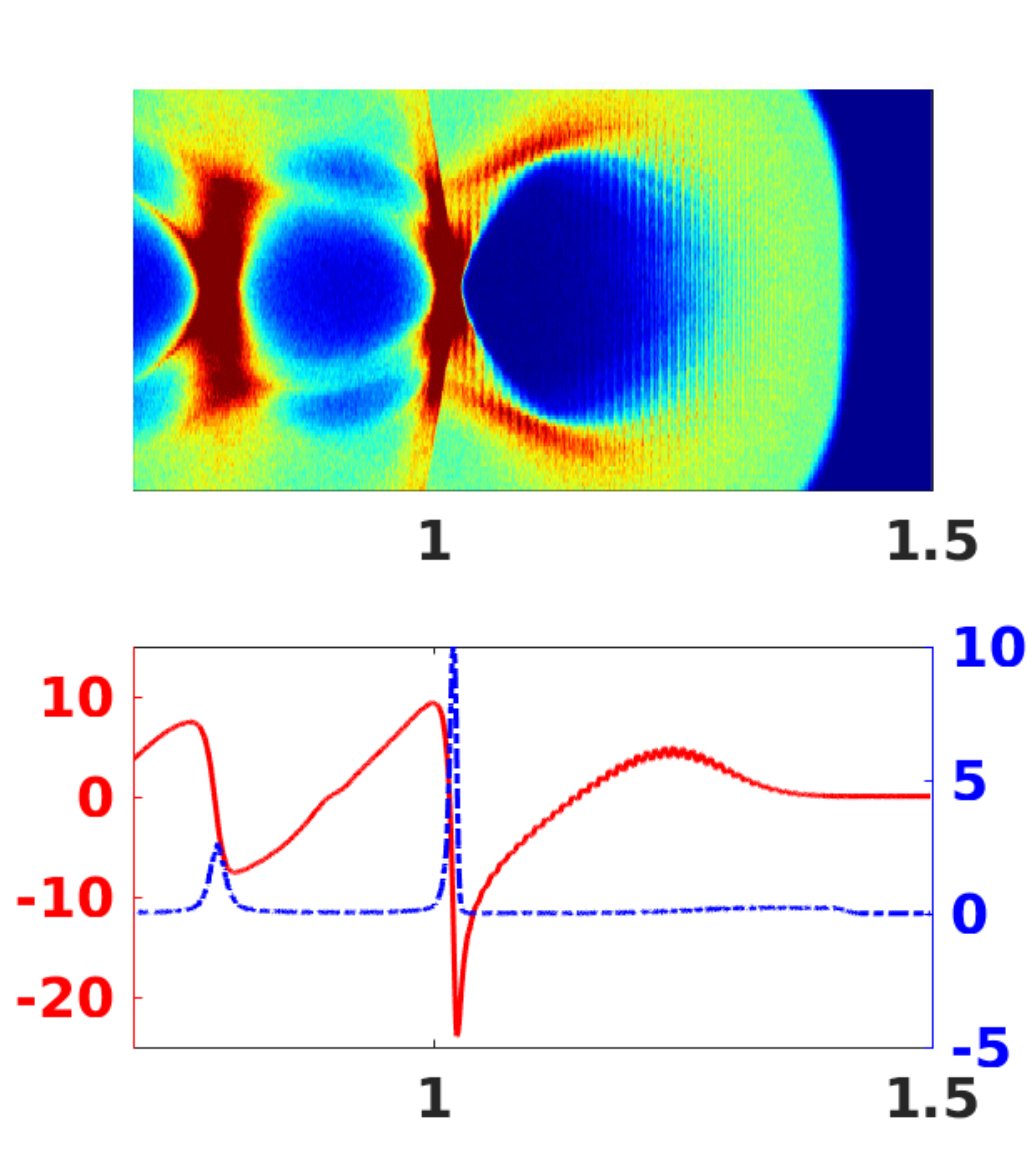}
    	\caption{ }
    	\label{b}
    \end{subfigure}	
    \begin{subfigure}{0.24\linewidth}
    	\includegraphics[width=\linewidth]{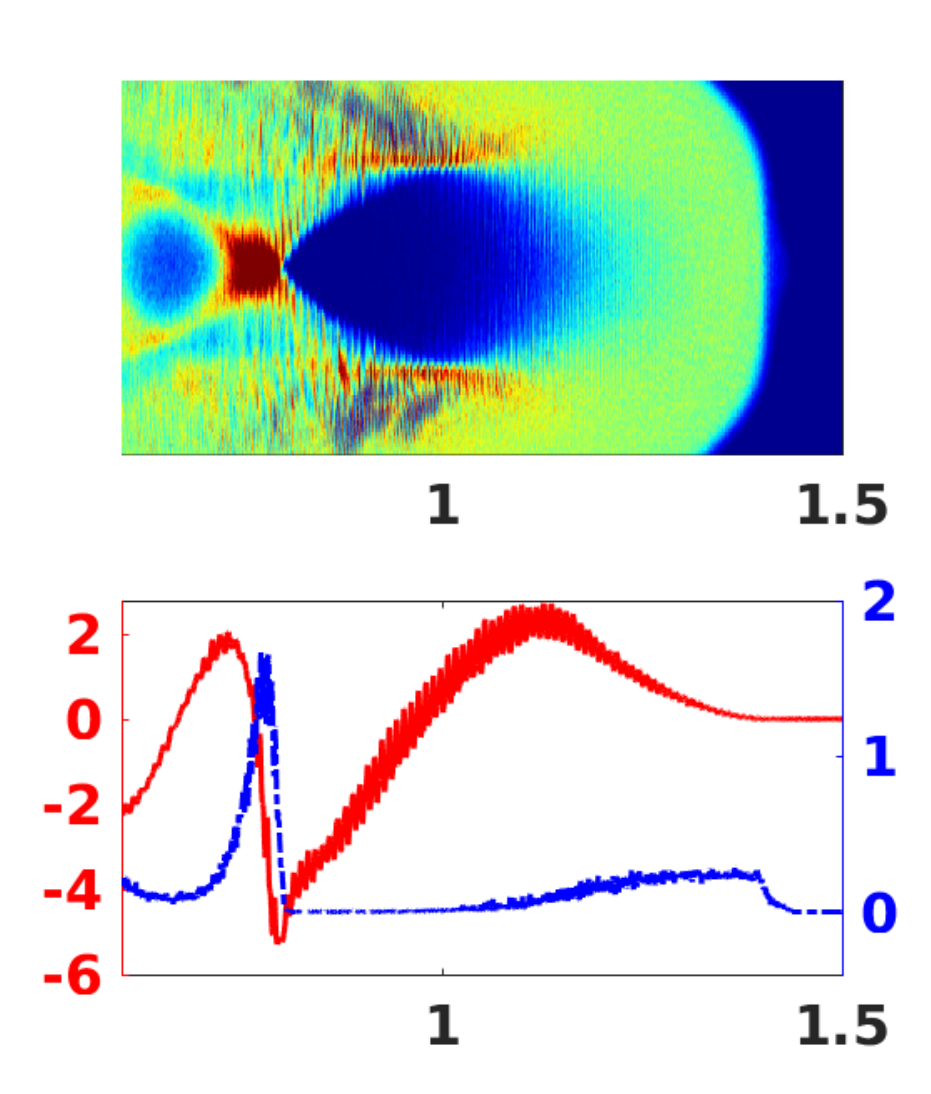}
    	\caption{ }
    	\label{c}
    \end{subfigure}	
    \begin{subfigure}{0.25\linewidth}
    	\includegraphics[width=5cm, height=5cm]{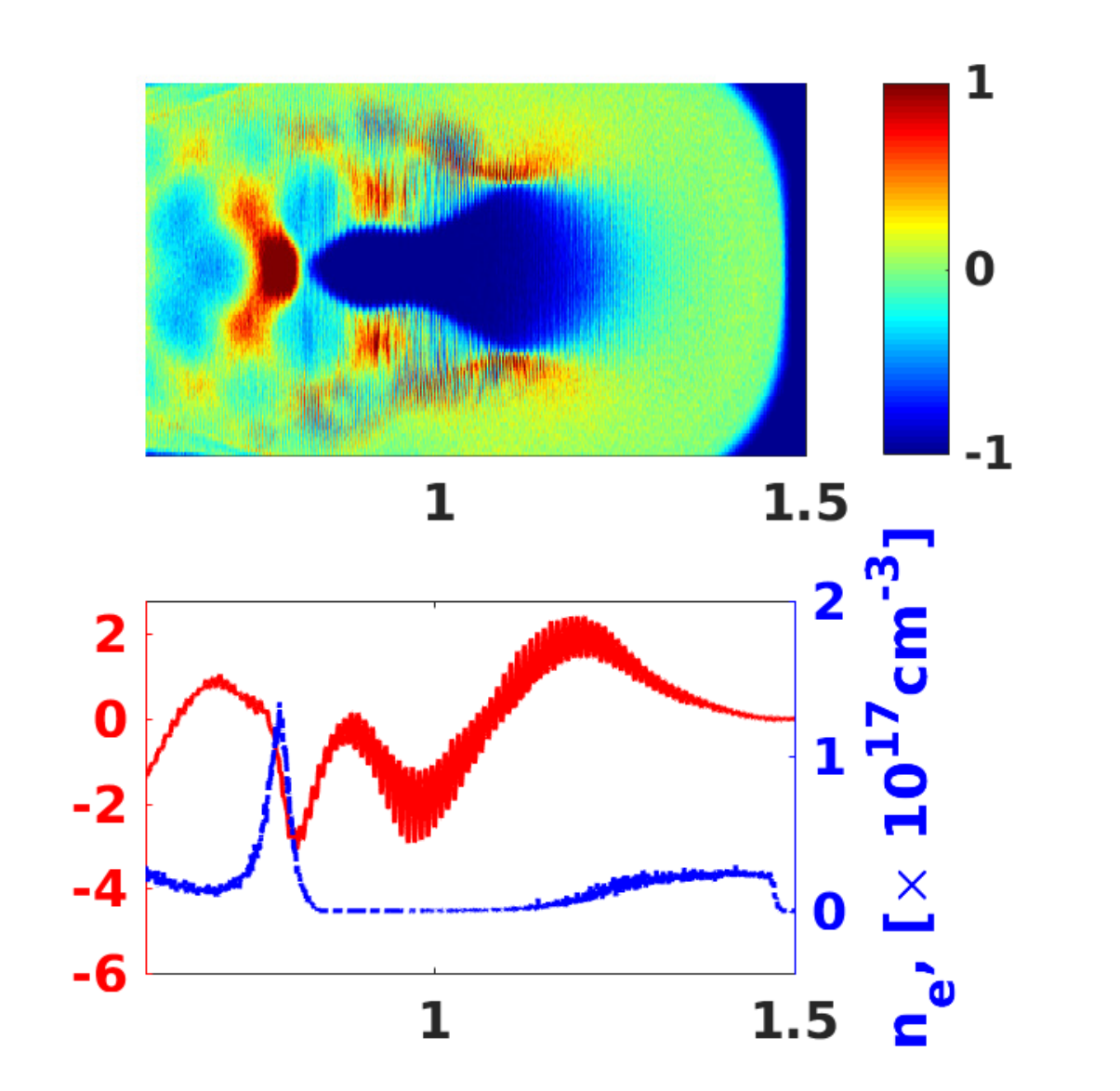}
    	\caption{ }
    	\label{d}
    \end{subfigure}	
	\caption{  \raggedright (a) $a_0 = 2.4: P = 20$ TW, $\tau = 0.5$ ps, $w_0 = 117$ $\mu m$, $n_e=2\times 10^{16}$ $cm^{-3}$. (b) $a_0=3.8: P=20$ TW, $\tau = 0.5$ ps, $w_0 = 74$ $\mu m$, $n_e = 2\times 10^{16}$ $cm^{-3}$. (c) $a_0 = 3.8 : P = 20$ TW, $\tau = 1$ ps, $w_0 = 74$ $\mu m$, $n_e = 2.3 \times 10^{16}$ $cm^{-3}$. (d) $a_0 = 4.0 : P = 10$ TW, $\tau = 1$ ps, $w_0 = 50$ $\mu m$, $n_e = 2.3\times 10^{16}$ $cm^{-3}$. Here $P: $ Laser Power, $\tau:$ pulse length, $w_0: $ spot size, $n_e : $ plasma density. Horizontal axis is the laser propagation direction in $mm$ and vertical axis on top row $(x)$ is the direction of linear polarization of the laser pulse. 
			}
	\label{blowout}	
\end{figure*}  
\subsection{Transition to blowout and bubble regimes}
High acceleration field and the linear focusing force of the blowout and bubble regime are
highly desirable for electron acceleration. LWIR drivers, which thanks to favorable wavelength
scaling are more efficient at driving plasma waves, can ultimately drive fully blown-out plasma bubbles with dimensions of several hundreds of microns. The large size of these bubbles relaxes the resolution requirements for the diagnostic probing of the properties of the wakefield as well as the alignment requirements for the external injection of an electron beam into the accelerating phase of the bubble. The latter is an important component of the experimental study of quality preservation in an LWFA in the blowout regime. To access this regime, the ponderomotive force of the laser has to be strong enough to expel virtually all of the electrons from a spherical region just behind the drive laser.

\begin{figure*}
	\centering
	\begin{subfigure}{0.25\linewidth}
		\includegraphics[width=4.9cm, height=5.6cm]{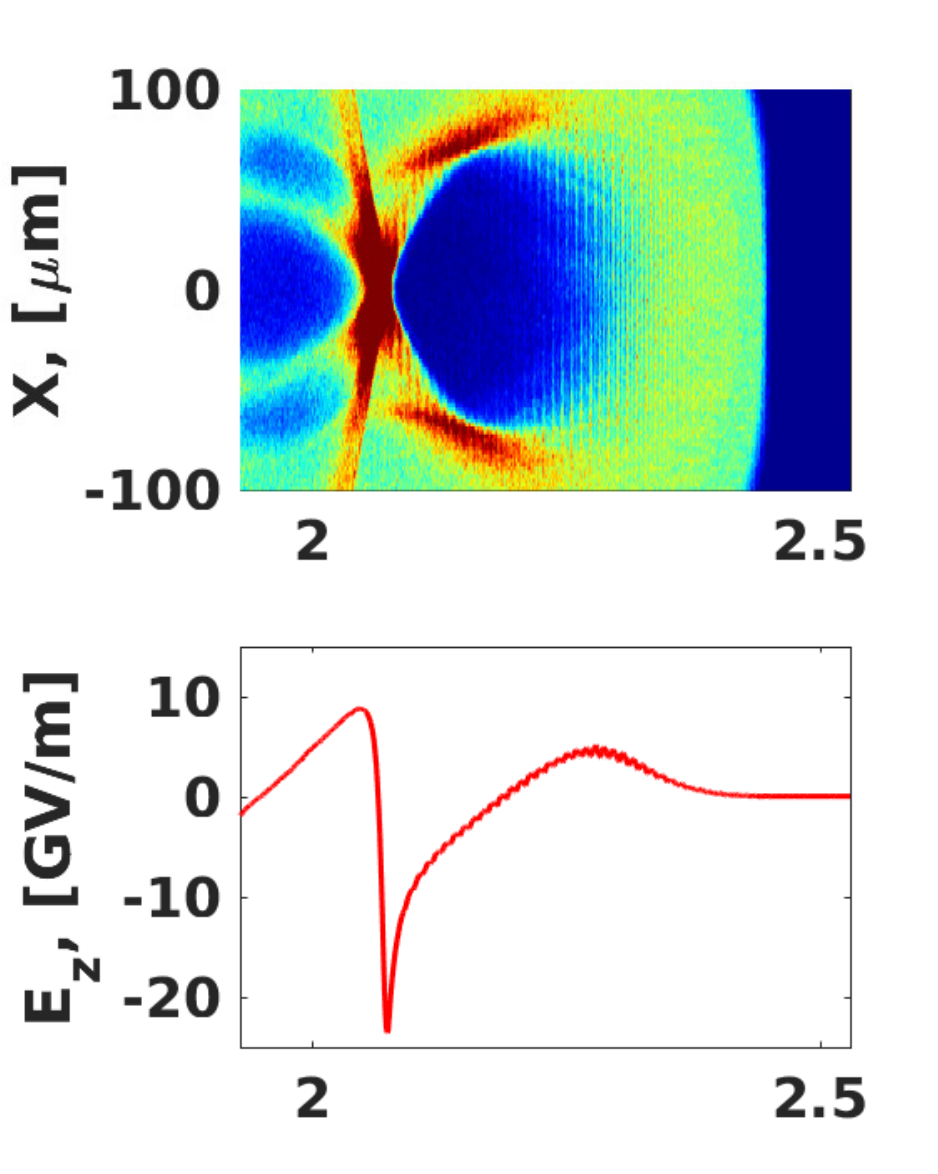}
		\caption{ $n_e = 2.0\times 10^{16}$ $cm^{-3}$}
		\label{a1}
	\end{subfigure}	
	\begin{subfigure}{0.24\linewidth}
		\includegraphics[width=\linewidth]{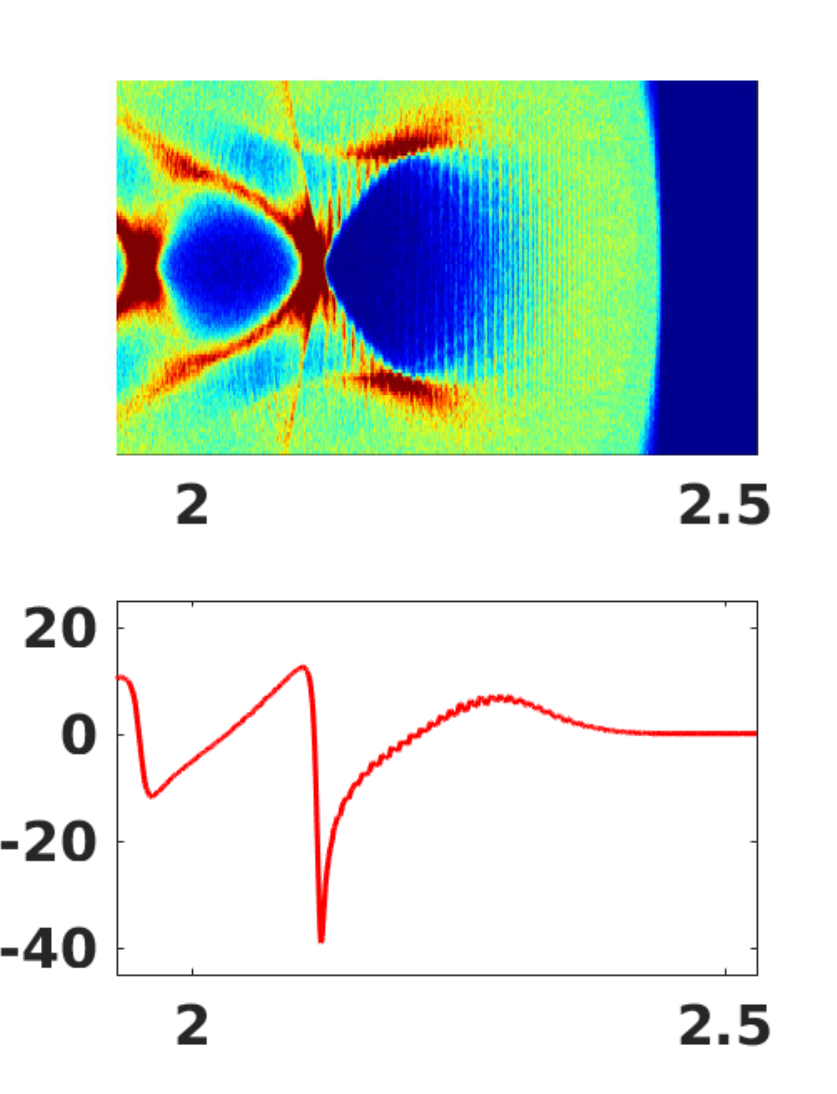}
		\caption{ $n_e = 4.0\times 10^{16}$ $cm^{-3}$}
		\label{b1}
	\end{subfigure}	
	\begin{subfigure}{0.24\linewidth}
		\includegraphics[width=\linewidth]{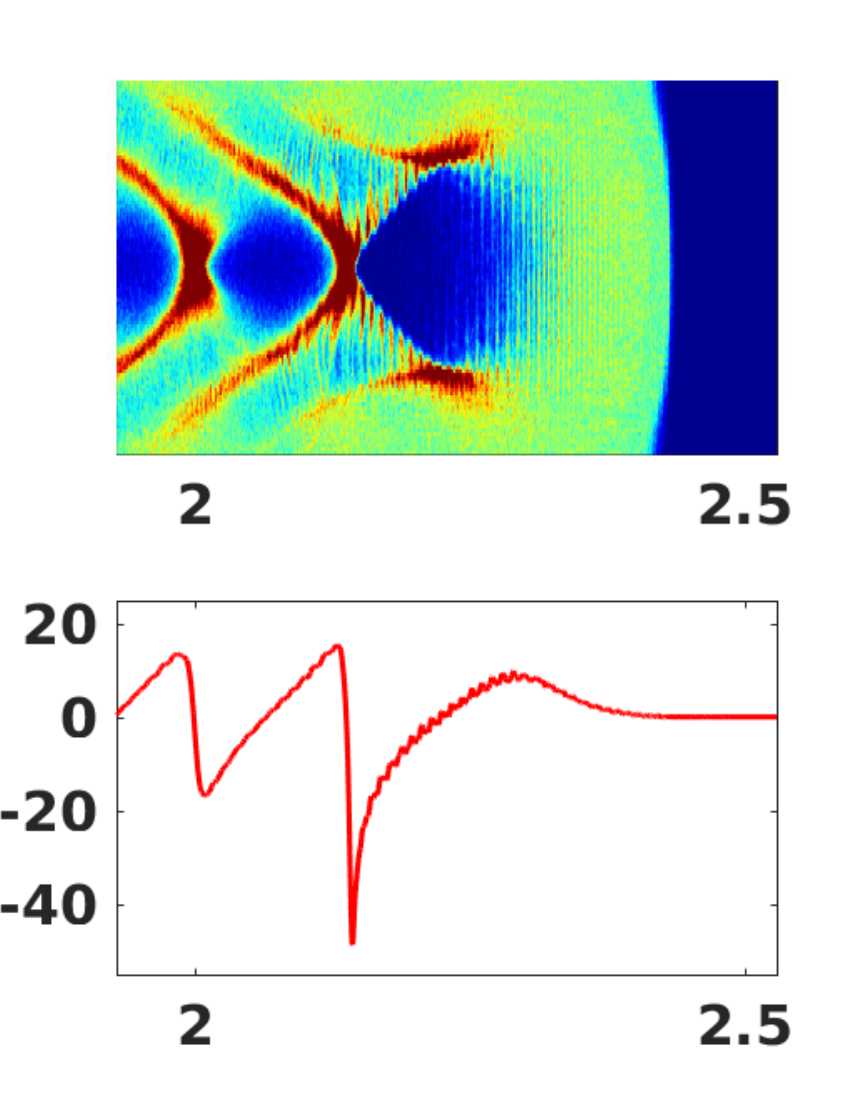}
		\caption{ $n_e = 6.0\times 10^{16}$ $cm^{-3}$}
		\label{c1}
	\end{subfigure}	
	\begin{subfigure}{0.25\linewidth}
		\includegraphics[width=5cm, height=5.6cm]{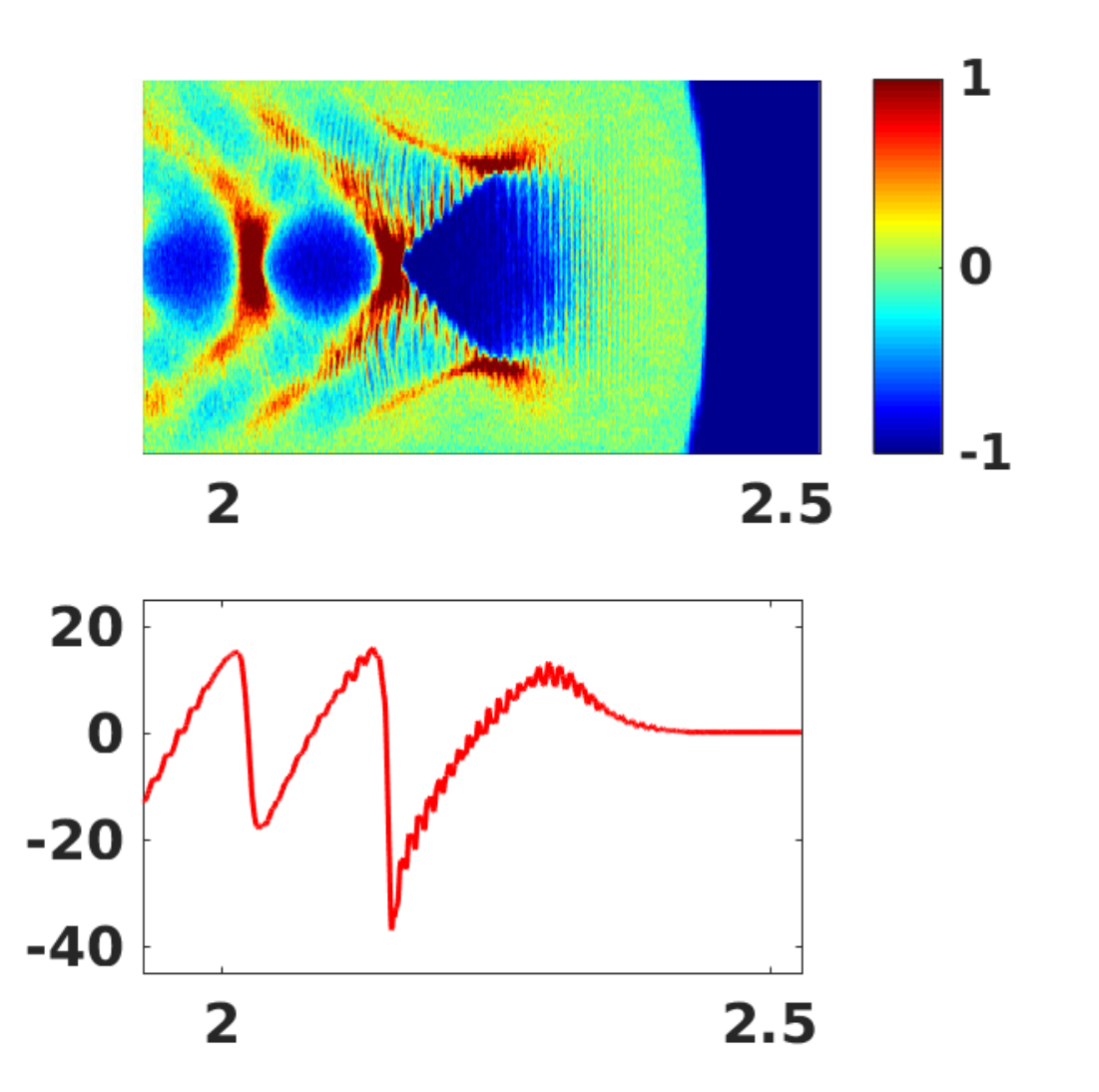}
		\caption{ $n_e = 8.0\times 10^{16}$ $cm^{-3}$}
		\label{d1}
	\end{subfigure}	
	\caption{  \raggedright Effect of variations in plasma density: Horizontal axis is the laser propagation direction in $mm$ and vertical axis on top row $(x)$ is the direction of linear polarization of the laser pulse. Higher longitudinal fields are produced at higher densities, due to stronger self-focusing of the laser beam, however, smaller plasma periods mean that laser pulse overlaps the accelerating phase of the wake for plasma densities greater than $ 4\times 10^{16}$ $cm^{-3}$ Laser $a_0 = 3.8: P = 20$ TW, $\tau = 0.5$ ps, $w_0 = 74$ $\mu m$.  
	}
	\label{blowout1}	
\end{figure*}  

Ideal performance is expected when the laser wakefield accelerator operates in the so-called “matched” mode, where the transverse and longitudinal dimensions of the laser pulse are approximately equal to the dimensions of the plasma wave it drives. Theoretical calculations based on non-linear theory \cite{Lu2006a} show that the transverse size of the laser is matched to the diameter of the bubble when $P/P_c = (a_0/2)^3$ . Here, $P_c = 17(\omega_0/\omega_p)^2$ is the critical power for self-focusing, $\omega_0$ is the laser frequency, $\omega_0$ is the plasma frequency, and $a_0 = 8.6\times 10^{-10}\sqrt{I[Wcm^{-2}](\lambda[\mu m])^2}$ is the normalized peak value of the vector potential of the laser. 

Longitudinally, a matched laser pulse will have a pulse length equal to the radius of the bauble. The ratio between the pulse length and the length of the nonlinear plasma wavelength can be characterized by a dimensionless pulse length parameter : $T_p = \frac{c\tau_{laser}}{\Lambda_{wake}} = \frac{\omega_p\tau_{laser}}{2\pi a_0^1/2}$. Here, $\tau_{laser}$ is the pulse length and $T_p = 1$ characterizes the condition where the laser fills the entire bubble. The matching condition therefore occurs for, which ensures that the accelerating electrons do not overlap with the laser fields before they reach their dephasing length, thus maintaining their transverse quality. The value of $a_0$ provides a good indicator for determining the regime of interaction. The ideal bubble regime (with a spherical ion column) is reached when $a_0$ reaches close to 4.

These matching criteria place stringent requirements on the laser parameters and can be used to estimate the threshold at which bubble regime operation will be possible (P > 27 TW for a $10$ $\mu m$ driver with a pulse duration of 0.5 ps). Alternatively, a “mismatched regime” can be used to increase the $a_0$ and access the bubble regime. For instance, the transverse matching condition for P = 20 TW requires a spot size of $117$ $\mu m$ for a plasma density of $2\times 10^{16}$ $cm^{-3}$, resulting in the matched $a_0$ value of 2.4. Because of the low value of $a_0$, the blowout regime is not reached in this configuration as shown in Figure \ref{a}. The blowout regime is reached by deviating from the matched condition by reducing the focus spot size such that the value of $a_0$ is raised to above three (a simulation with this is shown in Figure \ref{b}). 

Other simulations indicate that longitudinally mismatched laser pulses can also reach the blowout regime when $a_0 \sim 4$. Figure \ref{c} shows the result of a simulation also with a $P = 20$ TW, but with a pulse length of 1 ps. The spot size of 74 $\mu m $ results in an $a_0 = 3.8$ at a density of $2.3\times 10^{16}$ $cm^{-3}$, resulting in the blowout regime. Similarly, blowout regime is reached for a laser pulse with 10 TW peak power and 1 ps pulse length once the laser pulse is focused tightly (spot size of 50 $\mu m$) to reach an $a_0$ value of 4 (Figure \ref{d}). It is important to note however that this scaling does not hold if the pulse length is much longer than the plasma wavelength. For a 2 ps laser pulse at $7.5\times 10^{17}$ $cm^{-3}$ with $a_0 \sim 4$, the laser pulse is in the self-modulated regime, where the plasma bubble is disrupted after a few plasma periods. Therefore, in addition to an appropriately high  value, it is important for the laser pulse to be on the order of the bubble size, which for $\sim 2\times 10^{16}$ $cm^{-3}$, it means a pulse length less than 1 ps. 

The self-injection process to trap and accelerate electrons, studied in the previous sections, is sensitive to variations in laser and plasma parameters, and can be difficult to control and manipulate. Recent efforts in the field, therefore, have been focused on developing controlled injection mechanisms that do not rely on self injection, such as two-color ionization injection \cite{Yu2013}, down-ramp injection \cite{Barber2017}, or colliding pulse injection \cite{Malka2009}. For controlled injection methods to result in a high-quality electron beam, it is essential to suppress the self-injection, so that the quality of the injected electron beam is not affect by the addition of this “dark-current”. Suppressing the dark current will also be an important consideration in the external injection experiments, where the goal is to accelerate the injected bunch while preserving its transverse and longitudinal beam quality.

Previous simulation results had indicated that self-injection occurs when $a_0$ approaches 4
\cite{Lu2006a}, which is important because this value of $a_0$ is also required for
transitioning the LWFA into the blowout regime. In contrast, the simulations presented here
show clear evidence that LWIR lasers can drive a plasma wakefield in the blowout regime
with $a_0 \sim 4$ without triggering the process of self-injection. Figure \ref{a1} shows that a 0.5 ps laser pulse can drive moderately nonlinear plasma waves for $n_e = 2\times10^{16}$ $cm^{-3}$ producing wakefields of more than 20 $GV/m$ without any self-injection for a propagation distance up to $\sim 2.5$ $mm$. The relationship between the laser spot size and the matched spot size value of the LWIR laser pulse is likely responsible for reducing the transverse evolution of the blowout regime, which suppresses the self-injection of the electrons into the LWFA here. To examine whether different matching conditions would results in triggering of the self-injection process, several simulations were conducted with the same laser pulse as Figure \ref{a1}, but at gradually increasing plasma density up to $n_e = 8\times10^{16}$ $cm^{-3}$, with the results shown in Figures \ref{b1} - \ref{d1}. Importantly, self-injection was not observed in any of these cases. A second method of changing the laser transverse matching was examined by reducing the spot size by nearly half to 38 microns with plasma density at $n_e = 2\times10^{16}$ $cm^{-3}$. Even though in this case (not shown), the value of $a_0$ was significantly increased (to $a_0 = 7.4$), self injection was not observed. The absence of self-injected electrons makes the LWIR-driven blowout regime ideal for controlled injection experiments in an LWFA.  

Finally, the pulse length of the laser is an important experimental consideration for preserving the quality of an injected electron beam. For the simulation presented in Figure \ref{a1} for instance, the pulse length was 0.5 ps. At this pulse length, the laser pulse does not overlap the accelerating phase of the plasma wake. However, with increase in plasma density in Figures \ref{b1} - \ref{d1}, the wakefield amplitude increases, and the plasma wavelength and hence the bubble size gets smaller. This results in overlap of the laser pulse on longitudinal electric fields, which should be avoided to minimize distortions in the injected bunches.

\section{\label{Conclusions}Conclusions}
LWIR laser driven wakefield accelerators have been studied in the self-modulated and blowout regimes using 3D Particle-in-Cell simulations at parameters motivated by experiment AE-93 at ATF/BNL. Current laser parameters have allowed the exploration of self-modulated laser wakefield acceleration regime. For a linearly polarized laser pulse of duration $2.0$ ps and energy $4.0$ J, simulations show that background plasma electrons can get self-injected for plasma densities above $1\times 10^{17}$ $cm^{-3}$. Above this threshold density, the amount of injected charge and energy increases with increase in plasma density. Self-injection is shown to occur for wakefield amplitudes $\sim 24 \% $ of the theoretically predicted wave-breaking limit. Structure and angular distribution of accelerated bunches have been analyzed. Divergence, $px/pz$, where $x$ is the direction of linear polarization and $z$ is the direction of propagation of the laser pulse, shows a fork like structure. Simulations at higher energy (15 $J$) shows that the wave breaking occurs after only a few plasma wave periods, and a channel of approximately 80 $\mu m$ diameter is formed behind that. Electrons which are trapped in the channel gain forward momentum from the laser pulse through DLA process and get accelerated to multi-MeV energies. The fork like structure in the angular spectrum of the accelerated electrons is more prominent in this case.

Simulation predictions to transition from self-modulated regime to blowout regime has been presented. Blowout regime is reached for a laser pulse with $a_0 \sim 4$ for plasma density, $n_e \sim 2\times 10^{16}$ $cm^{-3}$. For a $1.0$ ps laser pulse this can be reached by a 20 TW laser focused to 74 $\mu m$ or a 10 TW laser focused to 50 $\mu m$ spot size. However, wakefield amplitude in these cases is weak and reaches only up to 4-6 $GV/m$. Moreover, laser pulse is long enough to overlap more than one plasma bubble producing distorted wakefields. Clean plasma bubble of more than 100 $\mu m $ diameter is produced by a 20 TW peak power laser with 0.5 ps pulse length for plasma density $n_e = 2\times 10^{16}$ $cm^{-3}$. Effect of variation in plasma density and increasing the propagation distance has been studied. At higher densities, plasma wavelength decreases, reducing the size of the bubbles. Although higher longitudinal fields are produced at higher densities, due to stronger self-focusing of the laser beam, smaller plasma periods mean that laser pulse overlaps the accelerating phase of the wake for plasma densities greater than $\sim 4\times 10^{16}$ $cm^{-3}$. The bubble produced for plasma density $n_e = 2\times 10^{16}$ $cm^{-3}$ with a 0.5 ps, 20 TW laser pulse generates accelerating fields as high as 24 $GV/m$, does not experience any self-injection for up to 2.5 mm long propagation, and can be used for external injection experiments.

\begin{acknowledgments}
This work was supported by the grant DE-SC0014043 funded by the U.S. Department of Energy, Office of Science, High Energy Physics. R. Z., M. C. D., and J. W. acknowledge additional support from AFOSR grant FA9550-16-1-0013 and DoE grant DE-SC0011617. Authors acknowledge support by staff of Accelerator Test Facility at Brookhaven National Laboratory.
\end{acknowledgments}

\bibliography{Kumar_lptb}

\end{document}